\documentclass[journal]{IEEEtran}

\usepackage{xcolor,soul,framed} 

\colorlet{shadecolor}{yellow}
\usepackage[pdftex]{graphicx}
\graphicspath{{../pdf/}{../jpeg/}}
\DeclareGraphicsExtensions{.pdf,.jpeg,.png}

\usepackage[cmex10]{amsmath}
\usepackage{amssymb}
\usepackage{bm}
\usepackage{array}
\usepackage{mdwmath}
\usepackage{mdwtab}
\usepackage{eqparbox}
\usepackage{url}
\usepackage{cite}
\usepackage{comment}
\usepackage[font=small]{caption}
\usepackage{tikz}
\usepackage{graphicx}
\usetikzlibrary{positioning}
\usepackage[labelformat=simple]{subcaption}
\usepackage{float}

\usepackage{siunitx}
\begin{document}
\bstctlcite{IEEEexample:BSTcontrol}
    \title{A TDD Distributed MIMO Testbed Using a 1-Bit Radio-Over-Fiber Fronthaul Architecture}
  \author{Lise Aabel,
      Sven Jacobsson,
      Mikael Coldrey,
      Frida Olofsson,\\
      Giuseppe Durisi,~\IEEEmembership{Senior Member,~IEEE,}
      and~Christian~Fager,~\IEEEmembership{Senior Member,~IEEE}

  \thanks{The work of L. Aabel, C. Fager, and G. Durisi was supported in part by the Swedish Foundation for Strategic
Research under Grant ID19-0036.}
  \thanks{L. Aabel is with Ericsson AB, 41756 Gothenburg, Sweden, and also with Chalmers University of Technology, 41296 Gothenburg, Sweden
(e-mail: lise.aabel@ericsson.com).}
  \thanks{S. Jacobsson and M. Coldrey are with Ericsson AB, 41756 Gothenburg, Sweden  (e-mail: sven.jacobsson@ericsson.com; mikael.coldrey@ericsson.com).}
  \thanks{F. Olofsson, G. Durisi and C. Fager are with Chalmers University of Technology, 41296 Gothenburg,
Sweden (e-mail: frida.olofsson@chalmers.se;  durisi@chalmers.se; christian.fager@chalmers.se).}
  \thanks{This paper was presented in part at the Asilomar Conference on Signals, Systems, and Computers, 2020 \cite{aabel_asilomar20}.}
}

\maketitle

\begin{abstract}

We present the uplink and downlink of a time-division duplex distributed multiple-input multiple-output (D-MIMO) testbed, based on a 1-bit radio-over-fiber architecture, which is low-cost and scalable.
The proposed architecture involves a central unit (CU) that is equipped with 1-bit digital-to-analog and analog-to-digital converters, operating at \SI{10}{GS/s}.
The CU is connected to multiple single-antenna remote radio heads (RRHs) via optical fibers, over which a binary RF waveform is transmitted.
In the uplink, a binary RF waveform is generated at the RRHs by a comparator, whose inputs are the received RF signal and a suitably designed dither signal.
In the downlink, a binary RF waveform is generated at the CU via bandpass sigma-delta modulation.
Our measurement results show that low error-vector magnitude (EVM) can be achieved in both the uplink and the downlink, despite 1-bit sampling at the CU.
Specifically, for point-to-point over-cable transmission between a single user equipment (UE) and a CU equipped with a single RRH, we report, for a \SI{10}{MBd} signal using single-carrier 16QAM modulation, an EVM of $\boldsymbol{3.3}$\% in the downlink, and of $\boldsymbol{4.5}$\% in the uplink.
We then consider a CU connected to 3 RRHs serving over the air 2 UEs, and show that, after over-the-air reciprocity calibration, a downlink zero-forcing precoder designed on the basis of uplink channel estimates at the CU, achieves an EVM of $\boldsymbol{6.4}$\% and $\boldsymbol{10.9}$\% at UE 1 and UE 2, respectively.
Finally, we investigate the ability of the proposed architecture to support orthogonal frequency-division multiplexing (OFDM) waveforms, and its robustness against both in-band and out-of-band interference.

\end{abstract}

\begin{IEEEkeywords}
Distributed MIMO, 1-bit sampling, radio-over-fiber.
\end{IEEEkeywords}

%
\IEEEpeerreviewmaketitle


\begin{table*}[t]
\caption{Uplink Measurements Using 1-Bit Radio-over-Fiber Fronthaul}
  \centering
    \begin{tabular}{ c|c c c c c c }
     \hline

     Ref. & Type & AGC & Dither gen. & Channel & Bandwidth & Fronthaul rate\\
    \hline
    \cite{aabel_asilomar20} & SISO & No & FPGA & Cable & \SI{10}{MHz} & \SI{10}{Gbps}\\

    \cite{Prata2016All-digitalC-RAN} & SISO & No & Sig. Gen. & Cable & \SI{5}{MHz} & NA \\
 
    \cite{IbraMe21} & SISO & No & FPGA & Cable & \SI{20}{MHz} & \SI{10}{Gbps} \\

    \cite{Maier2011ClassO} & SISO & No & NA & Cable & \SI{5}{MHz} & \SI{3}{Gbps} \\

    This work & Multi-user-D-MIMO & Yes & FPGA & Wireless & \SI{100}{MHz} & \SI{10}{Gbps} \\
 
     \hline
    \end{tabular}
    \label{tab:prevwork}
\end{table*}

\section{Introduction}

\IEEEPARstart{T}{he} demands for more uniform quality of service, lower latency, higher reliability, and improved energy efficiency are driving the wireless network design towards 6G \cite{ulf_6g21}.
Distributed multiple-input multiple-output (D-MIMO) is a promising technology to satisfy these demands, see, e.g.,  \cite{hexaX_23}.
In D-MIMO, a central unit (CU) is connected to multiple spatially separated antenna units, which we shall refer to as remote radio heads (RRHs).
The advantages of D-MIMO over traditional co-located MIMO include enhanced robustness towards shadow fading, reduced spatial correlation between channels, and increased energy efficiency \cite{zhou_dmimo03}.
From a network perspective, D-MIMO is a key technology to implement the cell-free massive MIMO network architecture described in, e.g., \cite{Ngo2017Cell-FreeCells,bjornson_cellfree,cellfreebook}. 

The theoretical principles underpinning the optimal design of D-MIMO architectures are well understood \cite{cellfreebook}.
However, many issues remain concerning the practical realization of this technology.
One challenge in the implementation of D-MIMO is to ensure phase-synchronized transmission and reception between the RRHs, which is necessary to serve simultaneously multiple user equipment (UEs) via spatial processing. 
In theoretical investigations, it is often assumed that frequency up- and down-conversion are performed locally at the RRHs, and that the fronthaul link connecting the RRHs to the CU carries samples of the baseband signals.
One way to achieve this is to equip each RRH with a local oscillator.
To guarantee phase-coherent transmission and reception, the local oscillators must then be synchronized.
This is challenging in the hardware domain, and costly in the digital domain, in terms of signaling overhead.

To bypass this issue, we consider in this paper an architecture in which up- and down-conversion are performed digitally at the CU.
This eliminates the issue of synchronizing local oscillators.
The drawbacks are that radio-frequency (RF) signals, rather than baseband signals, need to be exchanged over the fronthaul link and that analog-to-digital (ADC) and digital-to-analog (DAC) converters need to operate at a much higher sampling rate.
To alleviate both problems, we focus on an architecture in which the fronthaul links consist of optical fibers, and the RF signals are mapped into a two-level (binary) optical waveform, prior to transmission over the fiber.
In the following, we shall refer to this architecture as \emph{1-bit radio-over-fiber fronthaul}.
This architecture is advantageous in terms of cost, power consumption, and scalability \cite{FridaConf}.
Indeed, the use of binary optical waveforms allows us to leverage the recent development of low-cost, high-speed optical components, driven by the needs of, e.g., data centers.
Also, it allows us to use power-efficient 1-bit ADCs and DACs at the CU.

An alternative architecture based on analog radio-over-fiber fronthaul has also been considered in the literature for D-MIMO applications, see, e.g., \cite{puerta_arof}, although no phase-coherent downlink transmission based on uplink channel estimation has been demonstrated yet.
In this alternative architecture, the RF signal is modulated directly onto the optical carrier.
This, however, makes this architecture less robust toward nonlinearities occurring in the optical domain \cite{FridaConf}.

\subsection{Previous Work}\label{sec:prevwork}
D-MIMO architectures based on the 1-bit radio-over-fiber fronthaul have been studied previously in the literature \cite{aabel_asilomar20,Maier2011ClassO,Sezgin2018SISO,Sezgin2019MIMOtestbed,Sezgin2019MIMOeval,Wang2019,Wu_rfsdm20,Prata2016All-digitalC-RAN,IbraMe21}.
In the downlink, bandpass sigma-delta ($\Sigma \Delta$) modulation is used to map the RF signal into a two-level waveform \cite{Sezgin2018SISO,Sezgin2019MIMOtestbed,Sezgin2019MIMOeval,Wang2019,Wu_rfsdm20}.
At the CU, a 1-bit quantized and oversampled version of the RF signal is then transmitted to the RRHs via a two-level optical waveform.
The desired RF signal is finally obtained at each RRH via a bandpass filtering operation.
In the uplink, the RF signal is mapped into a two-level waveform at the RRHs by means of a comparator, which is fed with the RF signal and a suitably designed dither signal \cite{aabel_asilomar20,Prata2016All-digitalC-RAN,IbraMe21,Maier2011ClassO}.
The resulting two-level waveform is transmitted to the CU via the fiber optical link, where it is sampled with 1-bit resolution.

Bandpass $\Sigma\Delta$-over-fiber for D-MIMO fronthaul was demonstrated in \cite{Sezgin2018SISO,Wang2019} via point-to-point measurements.
In \cite{Sezgin2019MIMOtestbed,Sezgin2019MIMOeval,Wu_rfsdm20}, downlink transmission with multiple RRHs was demonstrated.
However, since only a downlink testbed was available, the channel-state information needed for the design of the precoder was transferred from the UEs to the CU via a cable. 
In \cite{Maier2011ClassO,aabel_asilomar20,Prata2016All-digitalC-RAN,IbraMe21}, point-to-point uplink reception was demonstrated.
The demonstrations involved only over-cable measurements, in which an RF signal produced by a signal generator was connected directly to one of the two ports of the comparator.
In \cite{Prata2016All-digitalC-RAN}, the dither signal was also produced by a signal generator, whereas in \cite{aabel_asilomar20,IbraMe21} it was generated at the CU and conveyed to the RRH via the downlink fronthaul.

A theoretical analysis of the 1-bit quantized uplink is presented in \cite{JacobssonGlobecom,Anzhong23}. 
In \cite{JacobssonGlobecom}, the authors show analytically and via simulations that the per-symbol error-vector magnitude (EVM) degrades in the high SNR regime when dithering is not used.
The EVM degradation is caused by a high correlation between the signal and the quantization noise.
This correlation can be mitigated via dithering, and satisfactory EVM performance at high SNR can be achieved.
Specifically, both white Gaussian noise and uniformly distributed binary signals were considered in \cite{JacobssonGlobecom} as possible dither signals.
In \cite{Anzhong23}, the authors quantify the minimum fronthaul rate (sampling rate) required, for a D-MIMO architecture with 1-bit radio-over-fiber fronthaul and a fixed number of RRHs, to outperform in terms of EVM a conventional co-located massive MIMO architecture. 
Furthermore, the paper sheds light on the optimal dithering power for the case of Gaussian dithering signal.

\subsection{Contributions}
We present a time-division duplex (TDD) D-MIMO testbed with 1-bit radio-over-fiber fronthaul and multiple RRHs.
Compared to previous work on 1-bit radio-over-fiber fronthaul for D-MIMO, our testbed implements both uplink and downlink, and supports downlink spatial multiplexing via reciprocity-based uplink channel estimation performed at the CU.
Compared to \cite{Maier2011ClassO,aabel_asilomar20,Prata2016All-digitalC-RAN,IbraMe21}, the RRH receiver is complemented with filters, a TDD switch, amplifiers, and automatic gain control (AGC).
In particular, the AGC provides a dynamic range to the RRH receiver, which allows us to apply the same optimized dither signal to received signals having different power.
We use point-to-point over-cable measurements to determine the dynamic range available in the uplink, the ability of the proposed architecture to support 5G waveforms, and also its robustness against both in-band and out-of-band interference.

We also present over-the-air measurements involving reciprocity calibration and downlink multi-user transmission based on uplink channel estimation, for a CU connected to $3$~RRHs serving $2$~UEs.
Our measurements reveal that low EVM can be achieved in both uplink and downlink, despite the 1-bit quantization, using quantization-unaware linear spatial processing at the CU.

In Table \ref{tab:prevwork}, we provide a summary of the uplink measurements conducted in \cite{Maier2011ClassO,aabel_asilomar20,Prata2016All-digitalC-RAN,IbraMe21} using the 1-bit radio-over-fiber fronthaul, as well as of the novel ones presented in this paper.

\subsection{Paper Outline}
The rest of this paper is organized as follows. 
In Section \ref{sec:trx_fun}, we detail the testbed architecture. 
In Section \ref{sec:trx_hw}, we discuss the testbed hardware and its operations. 
In Section \ref{sec:front_eval}, we present point-to-point over-cable measurements for downlink and uplink.
In Section \ref{sec:calibration}, we discuss system reciprocity calibration. 
We present over-the-air multi-user D-MIMO measurements in Section \ref{sec:dmimo}. 
Finally, we provide some concluding remarks in Section \ref{sec:concl}. 


\section{Testbed Architecture}\label{sec:trx_fun}

\begin{figure*}
    \centering
    \includegraphics[width=\textwidth]{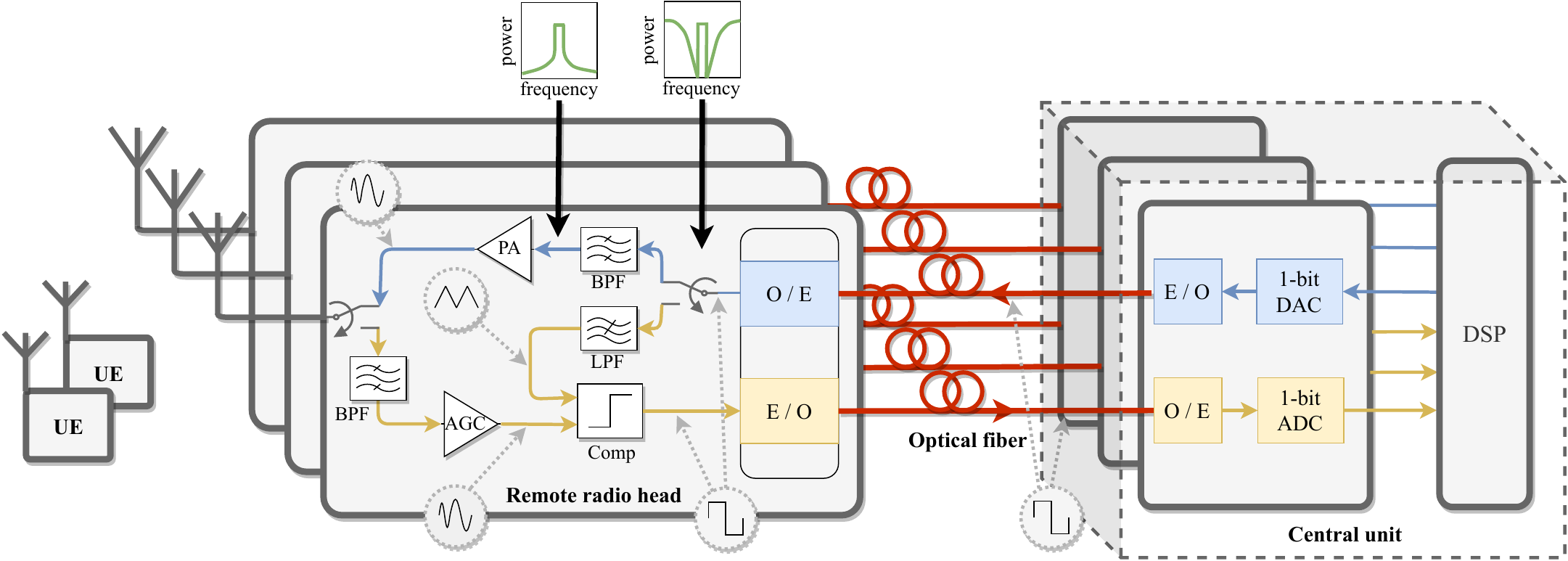}
    \caption{Block diagram of the proposed TDD D-MIMO testbed. The downlink (upper block chain in blue) consists of the following blocks: digital signal processing (DSP), 1-bit DAC, electrical-to-optical converter (E/O), optical fiber, optical-to-electrical converter (O/E), BPF, and PA. The uplink (lower block chain in yellow) consists of the following blocks: BPF, AGC, comparator (Comp), E/O, optical fiber, O/E, 1-bit ADC, and DSP. The dither signal in the uplink is conveyed to the RRHs via the downlink path, and reconstructed at the RRH by means of a LPF. The frequency spectrum of the signals in different parts of the architecture is illustrated by the green plots, and the waveforms are indicated by the dashed circles.}
    \label{fig:block_diagram}
\end{figure*}

A block diagram of the proposed D-MIMO testbed is provided in Fig.~\ref{fig:block_diagram}. 
In the figure, we consider the case in which a CU, connected to $3$~RRHs, serves $2$~UEs. 
The downlink and uplink are illustrated by the upper (marked in blue) and lower (marked in yellow) block chains, respectively. 
The signal generation and digital signal processing are performed at the CU, which is equipped with 1-bit DACs and ADCs, and is connected to the RRHs via optical fibers, carrying binary waveforms. 
In the next two subsections, we describe in detail the uplink and downlink blocks.

\subsection{Sigma-Delta-over-Fiber Downlink}\label{sec:dl}
At the CU, the coded information bits intended for each UE are first modulated and precoded.
Then, after pulse shaping, the baseband signal intended to each RRH is digitally up-converted to RF and then transformed into an oversampled 1-bit stream by a bandpass $\Sigma \Delta$ modulator.
This binary stream is converted to an analog signal by a 1-bit DAC, fed to an electrical-to-optical converter and sent to the corresponding RRH over a fiber optical cable.
At each RRH, the received optical binary signal is converted to the electrical domain by means of an optical-to-electrical converter.
Thanks to the properties of $\Sigma \Delta$ modulation, the RF signal is reconstructed at the RRHs simply by means of bandpass filtering.
Finally, the resulting signal is passed through a power amplifier (PA) and fed to the antenna port.

The principle of $\Sigma \Delta$ modulation is to use oversampling and noise shaping to place the quantization noise introduced by the 1-bit DAC outside the bandwidth of the desired signal.
As described in, e.g., \cite{Pavan_sdmBook17}, the amount of in-band quantization noise introduced by the $\Sigma \Delta$ modulator depends on two parameters:
1) the oversampling ratio

\begin{equation}\label{eqn:osr}
    \text{OSR} = \frac{f_s}{2W},
\end{equation}

\noindent
where $f_s$ is the sampling rate at which the bandpass $\Sigma \Delta$ modulator operates and $W$ is the bandwidth of the RF signal, and 2) the order of the $\Sigma \Delta$ modulator. 
Note that, the larger $f_s$, the lower the in-band quantization noise power. 
Furthermore, the order of the $\Sigma \Delta$ modulator, i.e., the number of integrators, determines the shape of the noise spectrum. 
The oversampling ratio and the order of the modulator can thus be adapted to achieve a desired in-band quantization noise power. 
However, higher-order modulators are more difficult to build, introduce more delay, consume more power, and are more prone to instability.
The parameters of the $\Sigma \Delta$ modulator implemented in the testbed are provided in Section \ref{sec:front_eval}.

\subsection{Dithered 1-bit Quantization Uplink}\label{sec:ul}
In the uplink, the RF signal at the antenna port of each RRH is first passed through a bandpass filter (BPF), and then through an AGC.
The AGC ensures that the amplitude of the filtered RF signal is within a desired range, to make dithering effective. 
A binary representation of the RF signal is then generated by feeding the output signal of the AGC and a dither signal to the comparator.
As demonstrated in, e.g., \cite{JacobssonGlobecom}, non-subtractive dithering can help whiten the quantization noise, which results in an improved EVM, especially in the high-SNR regime.

Similar to \cite{Prata2016All-digitalC-RAN,IbraMe21,Cordeiro2017,Mouton_pwm14}, we use a baseband triangular waveform as a dither signal.
In our architecture, the dither signal intended for each RRH is generated digitally at the CU.
This allows us to adapt the amplitude and the frequency of the dither signal to match the characteristics of the received RF signal.
This dither signal is then baseband $\Sigma \Delta$-modulated and conveyed to the RRHs via the downlink optical fibers, which are idle during the uplink in our TDD architecture. 
At the RRHs, the dither signal is reconstructed by means of a lowpass filter (LPF).
The binary output from the comparator, which resembles a pulse-width modulated RF signal, is provided to the electrical-to-optical converter and sent to the CU, where it is fed to an optical-to-electrical converter and sampled by a 1-bit ADC.
Digital signal processing in the uplink includes digital down-conversion, matched filtering, channel estimation, spatial combining, demodulation, and decoding.


\section{Hardware}\label{sec:trx_hw}
\begin{figure}
    \centering
    \includegraphics[width=\linewidth]{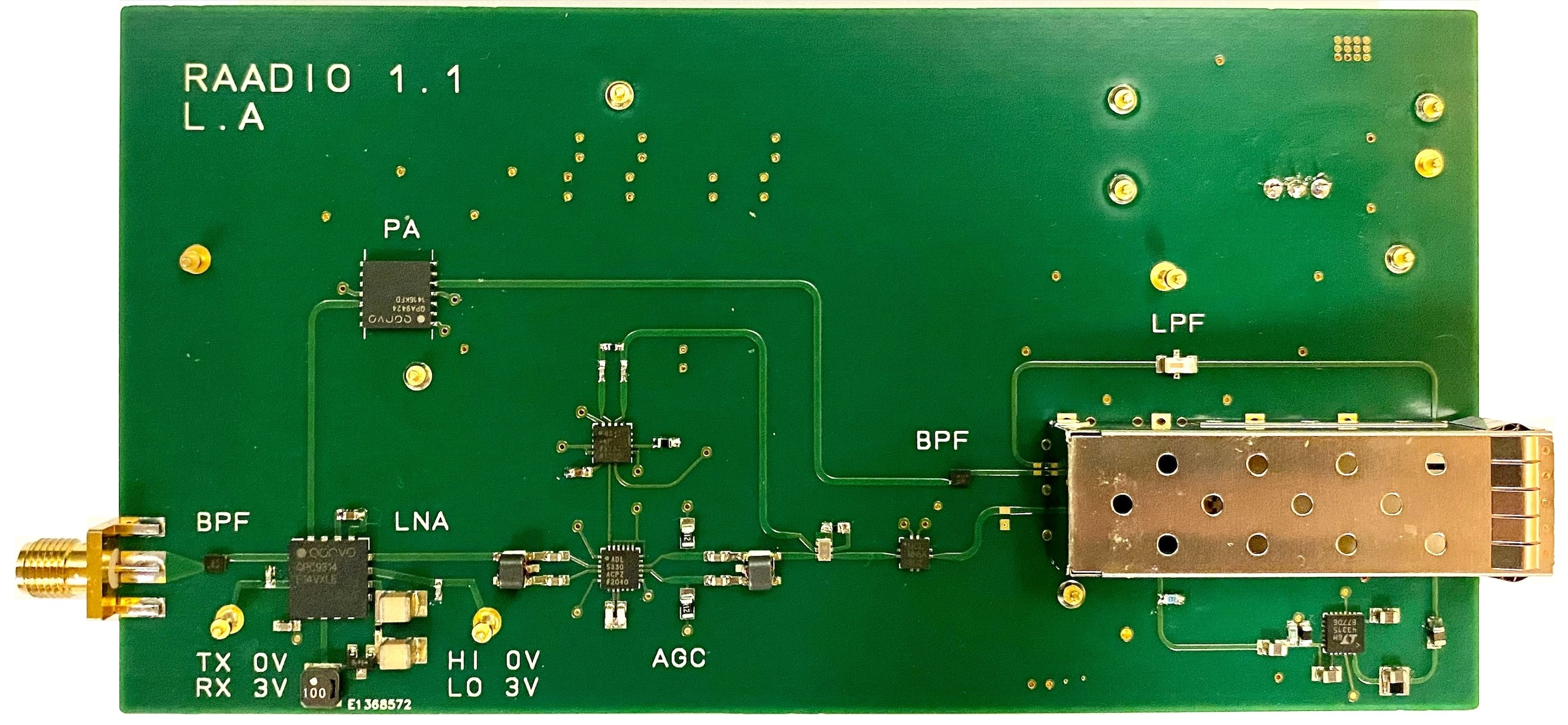}
    \caption{An RRH printed circuit board. The following components/functions are marked on the board: BPF, PA, LNA, AGC, and LPF. The settings for the switch/LNA are marked as TX and RX and the gain modes are marked as HI and LO.}
    \label{fig:front}
\end{figure}

\begin{figure}[H]
    \centering
    \includegraphics[width=0.95\linewidth]{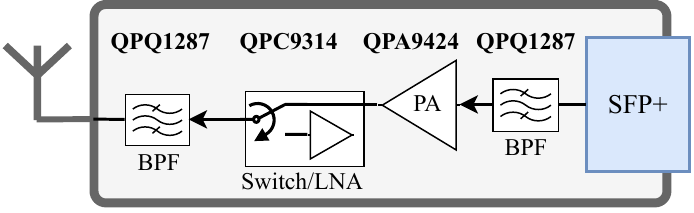}
    \caption{Block diagram of the RRH transmitter components. It includes two identical BPFs, a switch/LNA module, a PA, and an SFP+ optical transceiver.}
    \label{fig:dl}
\end{figure}

\begin{figure}
    \centering
    \includegraphics[width=0.95\linewidth]{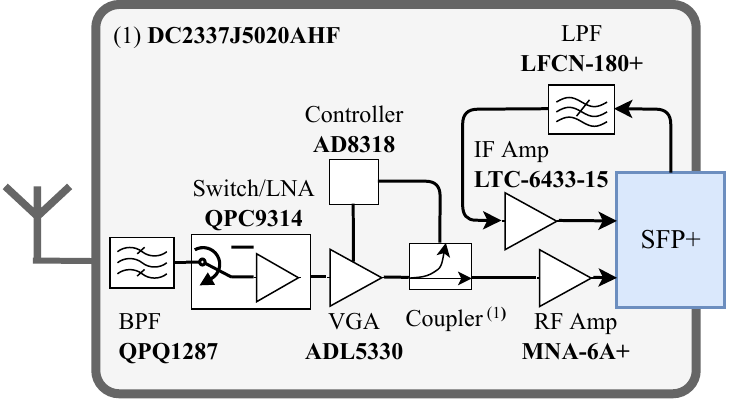}
    \caption{Block diagram of the RRH receiver components. It includes a BPF, a switch/LNA module, a VGA, a coupler, a demodulating logarithmic amplifier, an RF amplifier, a LPF, an intermediate frequency (IF) amplifier, and an SFP+.}
    \label{fig:ul}
\end{figure}

The D-MIMO testbed operates at a carrier frequency of \SI{2.35}{GHz}, which is dictated by the choice of the analog BPFs. 
Specifically, the BPFs used in the testbed are bulk acoustic wave reconstruction filters with \SI{100}{MHz} bandwidth, centered at \SI{2.35}{GHz}.
The CU consists of the field-programmable gate array (FPGA) evaluation board \textit{Altera Stratix V GT Transceiver Signal Integrity Development Kit} \cite{fpga} and a computer.
The FPGA evaluation board operates at a sampling rate of \SI{10}{GS/s}, and the digital ports are equipped with SMA connectors.
The digital signal processing is performed offline on the computer using MATLAB.
Specifically, $\Sigma \Delta$ modulation is implemented using the toolbox described in \cite{Pavan_sdmBook17}.
Electrical-to-optical and optical-to-electrical conversions are performed by the \textit{Avago AFBR-709SMZ} \textit{small form-factor pluggable+} (SFP+) optical transceivers \cite{sfp}. 
Each SFP+ is equipped with an \SI{850}{nm} vertical-cavity surface-emitting laser and photodetector, designed to support \SI{10}{Gigabit} Ethernet. 
The optical fiber cables in the testbed are \SI{30}{m} long, of type \textit{Optical Multimode 4}.
The electrical interface of the SFP+ offers differential input and output ports, which we use to implement the comparator. 
A picture of the assembled RRH is provided in Fig.~\ref{fig:front}.
We next describe the components of the transmitter, depicted in Fig.~\ref{fig:dl}, and the receiver, depicted in Fig.~\ref{fig:ul}, in the RRH.

\subsection{RRH Transmitter Components}
The downlink two-level RF signal from the CU is received at the RRH at the non-inverting electrical output port of the SFP+.
This signal is then passed through a reconstructing bulk acoustic wave BPF,\footnotemark{} centered at \SI{2.35}{GHz} and with \SI{100}{MHz} bandwidth.
The output of the filter is fed to a PA,\footnotemark[\value{footnote}] that has \SI{35.8}{dB} linear gain within the \SI{2.3}{}--\SI{2.4}{GHz} frequency range.
We use a switch connecting the antenna to either the transmitter or the receiver chain, to move between uplink and downlink operations.
Specifically, we use a combined switch/low-noise amplifier (LNA) module.
Switch synchronization is currently implemented in the testbed by applying a common DC power supply voltage to all RRHs.
When the switch is in the transmit state (downlink mode), the RF signal is connected directly to the output port of the module.
Before reaching the antenna port, the downlink signal is passed through another bulk acoustic wave BPF, which suppresses out-of-band emissions.

\footnotetext{This component is the same as the one used in \cite{Sezgin2019MIMOtestbed}.}

\subsection{RRH Receiver Components}\label{sec:hw-ul}
The RF signal received at the antenna port is filtered by the bulk acoustic wave BPF and connected to the switch/LNA module. 
When the switch is in the receive state (uplink mode), the RF signal is connected to the LNA within the module.
This amplifier operates within the \SI{2.3}{}--\SI{2.7}{GHz} frequency range, provides a high gain mode of \SI{33}{dB} or a low gain mode of \SI{24}{dB}, and has a noise figure of \SI{1.5}{dB}.
We use only the low gain mode during the measurements presented in this paper.
The AGC consists of a variable gain amplifier (VGA), which is controlled via a voltage applied to its gain control input port.
The VGA operates within the \SI{10}{MHz}--\SI{3}{GHz} range, provides a \SI{45}{dB} dynamic range (from \SI{-30}{dB} to \SI{15}{dB}), and has a noise figure of \SI{12.5}{dB}.
The voltage that controls the VGA is generated by a demodulating logarithmic amplifier, which acts as a power detector.
The demodulating logarithmic amplifier converts the RF signal at its input to a voltage that is proportional to the RF signal power in \SI{}{dB}. 
We use this demodulating logarithmic amplifier to maintain constant the average power of the signal at the output of the VGA at approximately \SI{-30}{dBm}.
To do so, we use a \SI{20}{dB} coupler to feed back a fraction of the RF signal at the output of the VGA to the demodulating logarithmic amplifier. 
The demodulating logarithmic amplifier is designed such that the adaptive controlling voltage does not change during \SI{1}{ms}.
After the AGC loop, i.e., at the output of the coupler, the RF signal is again amplified by an RF amplifier, to meet the input voltage requirements of the comparator.

The $\Sigma \Delta$-modulated triangular dither signal, sent from the CU via the fiber optical fronthaul, is received on the inverting output port of the SFP+ module and filtered by a \SI{180}{MHz} LPF to remove out-of-band distortions. 
The resulting triangular waveform is then amplified by a \SI{15}{dB} intermediate-frequency amplifier, operating within the \SI{100}{kHz}--\SI{1}{GHz}~range.

The output from the RF amplifier is fed to the SFP+ non-inverting input port and the output from the intermediate frequency amplifier is provided to the inverting input port.
The SFP+ thus serves both as comparator and electrical-to-optical converter.
Differential input voltages from \SI{180}{mV} to \SI{700}{mV} are supported by the SFP+.

\section{Point-to-Point Measurements} \label{sec:front_eval}
In this section, we present downlink and uplink over-cable point-to-point measurements. 
The purpose of these measurements is to investigate the RRH and fronthaul performance, without involving the wireless channel.
In all measurements, we use 16QAM symbols, root-raised cosine pulse shaping with a roll-off factor of $0.2$, and $4^{\text{th}}$ order $\Sigma \Delta$-modulation.
A $\Sigma \Delta$-modulator of order $4$ strikes a good tradeoff between noise-spectrum shaping and implementation complexity \cite{Pavan_sdmBook17}.

\subsection{Point-to-Point Downlink Over-Cable Measurements}
\begin{figure}[t]
    \centering
    \includegraphics[width=\linewidth]{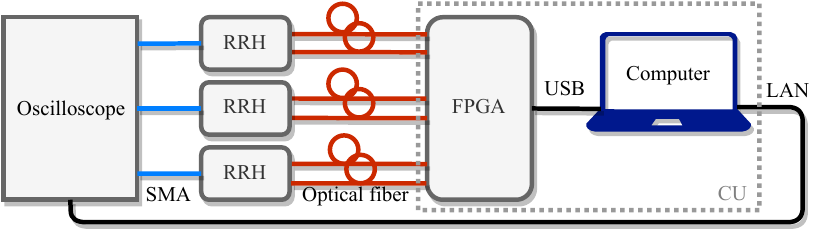}
    \caption{Setup for downlink over-cable measurements. The CU consists of the computer and the FPGA evaluation board, which are connected via a universal serial bus (USB). The oscilloscope has a \SI{4}{GHz} bandwidth and samples the signal at \SI{10}{GS/s}.}
    \label{fig:dl_lab_1}
\end{figure}

We measure the downlink performance for a \SI{10}{MBd} single-carrier RF signal in terms of EVM of the received constellation symbols, by using the setup described in Fig.~\ref{fig:dl_lab_1}. 
The antenna ports of the RRHs are connected to the input ports of the oscilloscope Rohde \& Schwarz \textit{RTO 1044}, which is connected to the computer via a LAN connection.
We then compare the EVM measured for this setup, with the one measured for the case in which the FPGA is replaced by a pulse-pattern generator (PPG).
A PPG provides high-quality binary waveforms with short rise time (\SI{8.5}{ps}).
Our goal with these two measurements is to separate the noise generated by the RRHs and the fronthaul link, from the noise generated by the FPGA evaluation board.

First, a single-carrier RF signal with \SI{10}{MBd} symbol rate is generated, bandpass $\Sigma \Delta$-modulated and written to the FPGA.
The binary waveforms at the output ports on the FPGA evaluation board are transmitted over the optical links to the RRHs.

\begin{figure}
\centering
\begin{subfigure}{\linewidth}
  \centering
  \includegraphics[width=\linewidth]{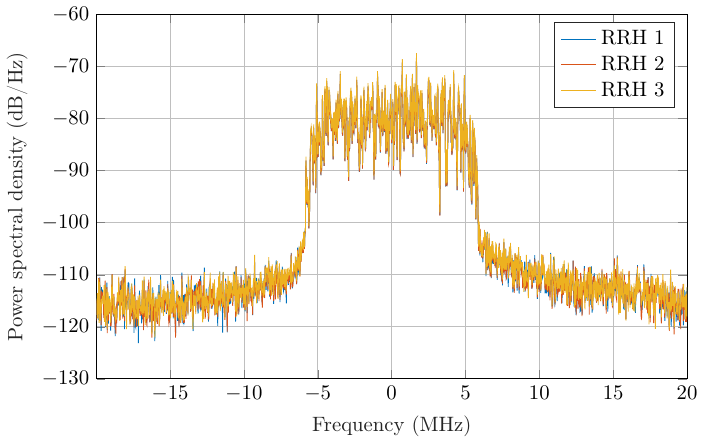}
  \caption{}
  \label{fig:aa}
\end{subfigure}
\begin{subfigure}{\linewidth}
  \centering
  \includegraphics[width=\linewidth]{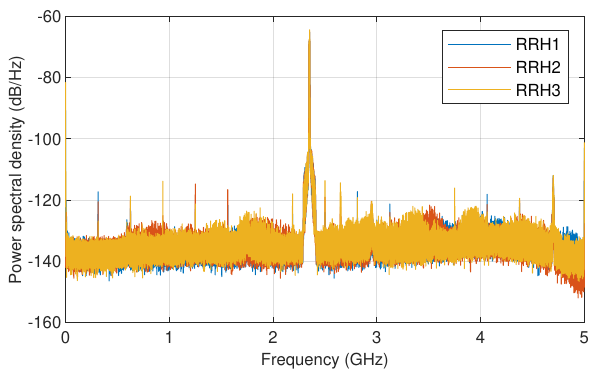}
  \caption{}
  \label{fig:ab}
\end{subfigure}
\par\bigskip 
\begin{subfigure}{.5\linewidth}
  \centering
  \includegraphics[height=0.8\linewidth]{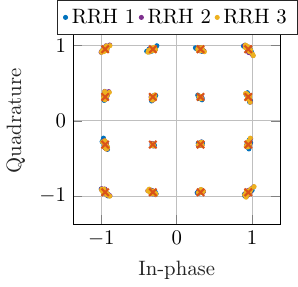}
  \caption{}
  \label{fig:bb}
\end{subfigure}%
\begin{subfigure}{.5\linewidth}
  \centering
  \includegraphics[height=0.8\linewidth]{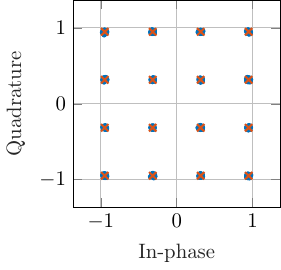}
  \caption{}
  \label{fig:cc}
\end{subfigure}

\caption{\subref{fig:aa} Welch power spectral density estimates of the in-band and adjacent bands when using the FPGA as binary waveform transmitter. \subref{fig:ab} Welch power spectral density estimates from \SI{0}{}--$f_s/2$ of the RF signals when using the FPGA as binary waveform transmitter. \subref{fig:bb} Constellation diagram obtained using 3 RRHs and FPGA. \subref{fig:cc} Constellation diagram obtained using RRH 1 and PPG.}
\label{fig:dl_cable}
\end{figure}

In Fig.~\ref{fig:aa} and \ref{fig:ab}, we present the baseband and passband power spectral densities of the signals received at the oscilloscope, when the same \SI{10}{MBd} signal is passed through each of the three RRHs.
We see from Fig.~\ref{fig:aa} that the power spectral densities are very similar.
Furthermore, the low noise outside of the bandwidth of interest seen in Fig.~\ref{fig:ab} indicates that the noise suppression of the BPFs is adequate.
We show in Fig.~\ref{fig:bb} the corresponding received constellation diagrams, which results in an EVM of $3.3\%$ for all RRHs.

\begin{figure}
\centering
\begin{subfigure}{\linewidth}
  \centering
  \includegraphics[width=\linewidth]{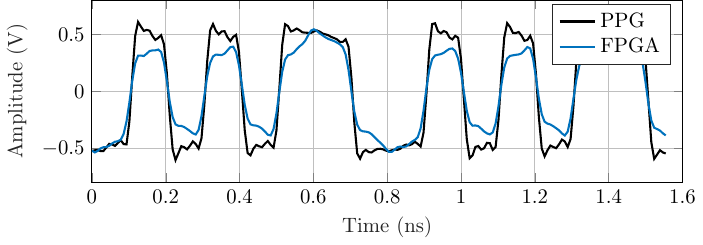}
  \caption{}
  \label{fig:dl_good}
\end{subfigure}
\par\bigskip 
\begin{subfigure}{\linewidth}
  \centering
  \includegraphics[width=\linewidth]{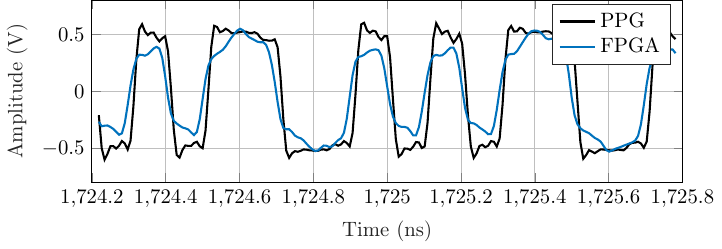}
  \caption{}
  \label{fig:dl_bad}
\end{subfigure}
\caption{Snapshot of FPGA and PPG output \subref{fig:dl_good} at the first time instance and \subref{fig:dl_bad} at the second time instance.}
\label{fig:dl_wvform}
\end{figure}

In Fig.~\ref{fig:cc}, we present the constellation diagram for the case in which the FPGA is replaced by the PPG and RRH $1$ is used.
The corresponding EVM is $2.3\%$. 
To understand the cause behind this EVM difference, we analyze next in detail the waveform generated by the FPGA and the PPG, corresponding to the bit pattern $[01010110]$ transmitted at \SI{10}{Gbps} and repeated for \SI{15}{$\mu$ s}.
The waveforms are recorded using a Keysight \textit{UXR0334A} oscilloscope with a bandwidth of \SI{33}{GHz}, which samples them at \SI{128}{GS/s}.
In Fig.~\ref{fig:dl_wvform} we provide a comparison between two snapshots of the two waveforms, taken at two different time instances.
We see from Fig.~\ref{fig:dl_good}, that, as expected, the pulses generated by the PPG have faster rise time.
As a consequence, the short pulses generated by the FPGA have lower amplitude than the ones generated by the PPG.
In Fig.~\ref{fig:dl_bad}, we present a second snapshot, recorded \SI{1724}{ns} after the first snapshot.
Compared to the first snapshot, we see that the FPGA signal is now also affected by an additional delay, which hints at a phase instability of the FPGA output. 
This phase instability as well as the amplitude reduction causes the additional noise observed in the constellation diagram in Fig.~\ref{fig:bb}, compared to Fig.~\ref{fig:cc}.

\subsection{Point-to-Point Uplink Over-Cable Measurements}\label{sec:ulp2p}
\begin{figure}[!t]
    \centering
    \includegraphics[width=\linewidth]{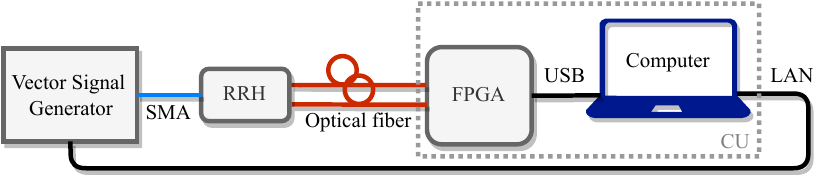}
    \caption{Setup for uplink over-cable measurements.}
    \label{fig:ul_labb_1}
\end{figure}

To measure the uplink performance in terms of the EVM of the received constellation symbols, we connect a Rohde \& Schwarz \textit{SMU 200A} vector signal generator via SMA cable to the antenna port of the RRH. 
The baseband signal is generated in MATLAB and provided to the vector signal generator, which performs frequency up-conversion. 
The measurement setup is shown in Fig.~\ref{fig:ul_labb_1}.

\subsubsection{Dithering Signal} 
We investigate the effect on the EVM of changing the power and the frequency of the dither signal, when feeding the RRH with a \SI{10}{MBd} signal.
During this measurement, the average transmitted power is set to \SI{-40}{dBm}.

\begin{figure}
    \centering
    \includegraphics[width=\linewidth]{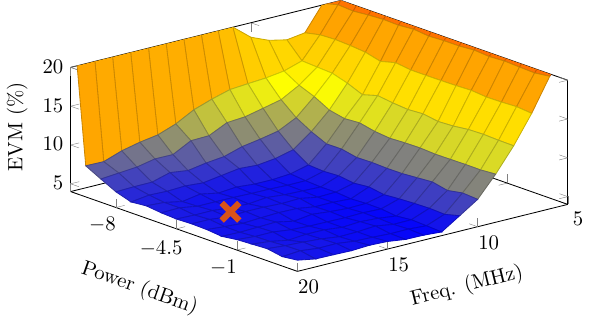}
    \caption{EVM of the received symbol constellation as a function of the frequency and the power of the dither signal. The minimum EVM, marked in the plot with a cross, is $4.5\%$. It is achieved for a dither signal power of \SI{-4.5}{dBm} and frequency of \SI{17}{MHz}.}
    \label{fig:dither}
\end{figure}

In Fig.~\ref{fig:dither}, we present the EVM of the received constellation symbols, as a function of the frequency and the power of the dither signal.
We observe that when the frequency of the dither signal is lower than the symbol rate of the transmit signal (i.e., \SI{10}{MBd}) or when the power of the dither signal is below \SI{-8}{dBm}, the EVM is high.
These results are in accordance with the ones reported in \cite{Cordeiro2017,IbraMe21}.
Furthermore, we observe that changing the dither signal frequency and power has little effect on the EVM when the frequency is higher than \SI{10}{MHz} and the power is higher than \SI{-8}{dBm}.
The smallest measured EVM is $4.5\%$ and occurs at \SI{17}{MHz} and \SI{-4.5}{dBm}, but many combinations of dither signal frequency and power yield similar performance.

\begin{figure}
\centering
\begin{subfigure}{.5\linewidth}
  \centering
  \includegraphics[height=0.8\linewidth]{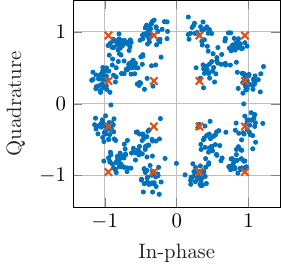}
  \caption{}
  \label{fig:ul_a}
\end{subfigure}%
\begin{subfigure}{.5\linewidth}
  \centering
  \includegraphics[height=0.8\linewidth]{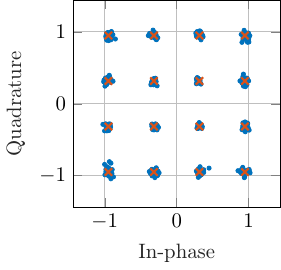}
  \caption{}
  \label{fig:ul_b}
\end{subfigure}
\caption{The constellation diagram in \subref{fig:ul_a} is obtained using a \SI{5}{MHz} dither signal at \SI{-11.5}{dBm} and in \subref{fig:ul_b} using a \SI{17}{MHz} dither signal at \SI{-4.5}{dBm}.}
\label{fig:ul_const}
\end{figure}

To illustrate the cause of the high EVM experienced when the choice of the power and frequency of the dither signal are suboptimal, we present in Fig.~\ref{fig:ul_a} the received constellation diagram for a poor choice of dither-signal power and frequency.
We observe from the figure that the signal is severely distorted by the 1-bit quantization.
In comparison, we illustrate in Fig.~\ref{fig:ul_b} the constellation diagram for a good choice of dither-signal power and frequency.
We note that the EVM is greatly improved and that, differently from Fig.~\ref{fig:ul_a}, the noise exhibits a circular symmetry around each constellation point. 


\subsubsection{Dynamic Range}
\label{sec:dynrange}
We investigate the dynamic range of the RRHs by measuring the EVM of the received constellation symbols for different values of the transmit signal power and a fixed dither signal. 
The measurements are conducted using the same signal parameters as for the dither analysis.
Furthermore, the dither frequency is set to \SI{16}{MHz} and the power to \SI{-4.5}{dBm}.

\begin{figure}
    \centering
    \includegraphics[width=\linewidth]{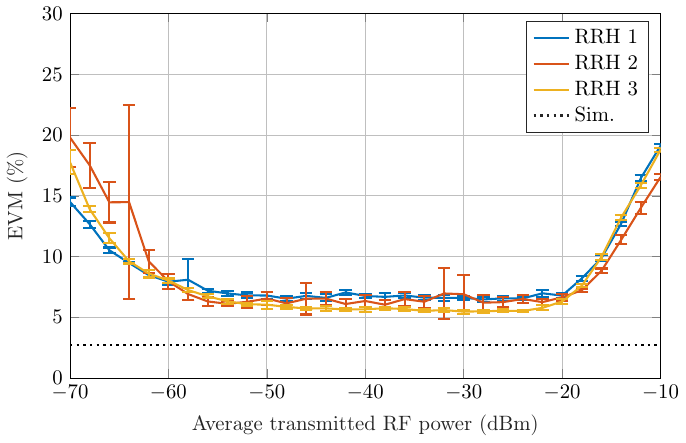}
    \caption{Average per-symbol EVM for each RRH, versus the average transmitted RF power from the vector signal generator. For each point, we provide the standard deviation, based on ten measurements. The simulated EVM, obtained by assuming ideal hardware components, is marked with a dashed line.}
    \label{fig:dynrange}
\end{figure}

In Fig.~\ref{fig:dynrange}, we report the measured EVM as a function of the average transmitted RF signal power. 
The error bar at each measurement point marks the standard deviation, estimated based on ten measurements.
From the figure, we obtain the following insights:
the flatness of the EVM curves in the interval [\SI{-60}{dBm}, \SI{-18}{dBm}] shows that the AGC keeps the average RF signal power constant within \SI{42}{dB}, which is close to the declared dynamic range of the VGA (\SI{45}{dB}).
The VGA has a constant noise figure; hence, the variable gain causes a change in the SNR of the signal.
However, we see from the figure that the EVM is approximately constant over the dynamic range of the AGC, which implies that this SNR change is negligible.
At the knee appearing just around \SI{-18}{dBm}, the VGA operates at \SI{-30}{dB} gain.
For a transmitted power higher than \SI{-18}{dBm}, the VGA operates non-linearly and the signal-to-dither power ratio is sub-optimal. 
At around \SI{-60}{dBm}, the VGA operates at \SI{15}{dB} gain.
For a transmitted power lower than this, the resulting low SNR, the VGA non-linearity, and the non-ideal dither signal properties contribute to the degradation in EVM. 

In the figure, we include also a simulated EVM of $2.7\%$, obtained by assuming ideal quantization, $\Sigma \Delta$-modulated and lowpass filtered dither signal, and additive white Gaussian noise with power equal to \SI{-26.6}{dBm} prior to dithering.
This value of noise power is obtained by assuming a thermal noise of \SI{290}{K} at the RRH antenna port, and using the noise figures, the gain of the amplifiers, and the bandwidths reported in the component data sheets to calculate the corresponding noise power.
Specifically, we use for the LNA, VGA and the RF amplifier the noise figures of \SI{1.5}{}, \SI{12.5}{} and \SI{2.7}{dB}, respectively.
This noise model is used in the remaining simulations of the receiver.
The difference between the simulated and measured EVM within the dynamic range reveals that the noise caused by the amplifiers in the RRH is not the only source of noise in the system.
Additional noise may be caused by the optical transceiver components as well as by non-ideal quantization and sampling in the CU.
This is discussed further in Section~\ref{sec:qerror}.

\subsubsection{Sampling and Quantization Errors at the CU}\label{sec:qerror}
\begin{figure}
    \centering
    \includegraphics[width=\linewidth]{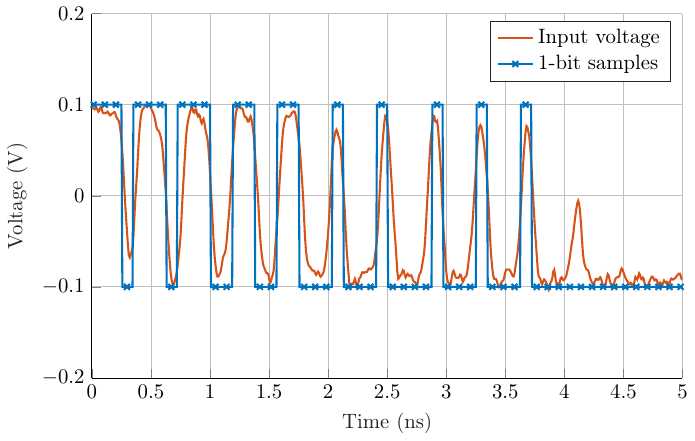}
    \caption{Input voltage to the FPGA digital port and 1-bit samples obtained by applying the sign-function.}
    \label{fig:ul_wvfrm}
\end{figure}

As shown in Fig.~\ref{fig:dynrange}, additional noise sources cause a degradation between $3\%$ and $4\%$ of the EVM.
To investigate potential errors introduced during quantization in the SFP+, we connect the output of the SFP+ at the CU to a Keysight \textit{UXR0334A} oscilloscope operating at \SI{128}{GS/s}, and measure the signal voltage.
The result is illustrated in Fig.~\ref{fig:ul_wvfrm}.
We start by noting that the input voltage to the FPGA is the output of the comparator at the RRH, after it is passed through the SFP+ at the RRH, the fiber-optical cable, and the SFP+ at the CU.
This waveform should ideally be a two-level waveform consisting of pulses of essentially arbitrary width.
Indeed, the width of a pulse is determined by the distance between two consecutive zero crossings of the signal obtained by subtracting the dither signal from the filtered RF signal received at the RRH.\footnotemark{}
The actual input voltage is a low-pass filtered version of a two-level waveform.
We measure an average rise and fall-time of \SI{54}{ps} of the input voltage pulses.
Hence, if the separation between two zero crossings is less than \SI{54}{ps}, the input to the FPGA will be significantly distorted.
Such a distorted pulse is observed in Fig.~\ref{fig:ul_wvfrm} at around \SI{4}{ns}, which may cause the FPGA decision circuitry to generate a bit at random.
We assessed the impact of the FPGA sampling by applying in the digital domain the sign function to the recorded input voltage to the FPGA and by using the corresponding quantized signal to perform demodulation.
This signal is illustrated by the blue curve in Fig.~\ref{fig:ul_wvfrm}.
This resulted in an EVM improvement of about $1\%$, compared to when the signal is sampled by the FPGA, which partly explains the discrepancy of $3\%$--$4\%$ between simulated and measured EVM.

\footnotetext{Note that the FPGA sampling rate is \SI{10}{GS/s}. Hence, the minimum duration of the pulses of the 1-bit waveform in Fig.~\ref{fig:ul_wvfrm} is \SI{0.1}{ns}.}

\subsubsection{Bandwidth}
\label{sec:bw}
The bandwidth of the receiver is determined by the bandwidth of the analog components in the RRH.
The BPFs have a bandwidth of \SI{100}{MHz} (\SI{2.3}{}--\SI{2.4}{GHz}), and the lower cutoff frequency of the LNA is \SI{2.3}{GHz}.
A reasonable assumption is that the receiver RF-chain in the RRH will not have a flat frequency response for input signals whose bandwidth is close to the bandwidth of the RRH components.
We investigate the bandwidth limitation of the RRH by presenting the EVM obtainable when receiving signals with symbol rate $R_s \in \{10,20,30,40,50,60,70,80,90\}$~\SI{}{MBd}.
For the case of root-raised cosine pulse shaped single-carrier signal with a roll-off factor of $\alpha = 0.2$, the corresponding bandwidth, computed as $R_s(1+\alpha)$, is $\{12,24,36,48,60,72,84,96,108\}$~\SI{}{MHz}, respectively.
Since the Rohde \& Schwarz \textit{SMU 200A} vector signal generator we used for the measurements in Section~\ref{sec:ulp2p} can generate signals with a maximum bandwidth of \SI{50}{MHz}, we use instead the Agilent Technologies \textit{M9502A} arbitrary waveform generator.
The average power of the input signal and of the dither signal are set to \SI{-30}{dBm} and \SI{-4.5}{dBm}, respectively, and we use the dither frequencies $\{16,22,31,37,45,61,69,77,81\}$ \SI{}{MHz}, which are found via an exhaustive search of the lowest EVM.
Furthermore, we compensate for possible inter-symbol interference caused by the hardware by passing the received signal through an $L$-tap equalizer, before demodulation.
The equalizer weights are obtained by least-square channel estimation.

\begin{figure}
    \centering
    \includegraphics[width=\linewidth]{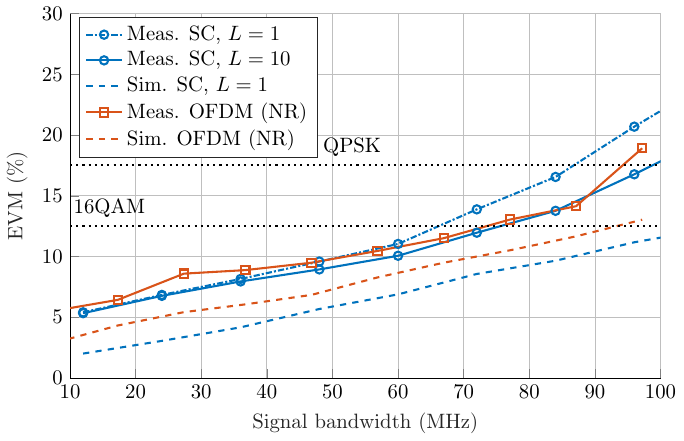}
    \caption{Average per-symbol EVM for different signal bandwidths. The figure presents measured and simulated results for single-carrier (SC) and OFDM waveforms.}
    \label{fig:bandwidth}
\end{figure}

In Fig.~\ref{fig:bandwidth}, we present the measured EVM for single-carrier modulation with single- and ten-tap equalization, and the corresponding simulated EVM.\footnotemark{}
In the simulations, we assume the receiver response to be frequency flat.
As shown in the figure, for the scenario considered in this section, the EVM increases as a function of the signal bandwidth.
For the case $L=1$, we see that this increase is roughly linear up to \SI{60}{MHz}, after which the EVM appears to increase faster than linear.
For the case $L=10$, the linear increase of the EVM holds up to \SI{84}{MHz}.
Furthermore, the resulting EVM values are lower than for the case $L=1$.
This confirms that the frequency response of the RRH is not flat for bandwidths larger than \SI{60}{MHz}.
According to the long-term evolution (LTE) and new radio (NR) standards, the minimum EVM requirement for 16QAM orthogonal frequency-division multiplexing (OFDM) signals is $12.5\%$, and it is $17.5\%$ for QPSK \cite{3gpp,5g}.
Although we use single-carrier signals in this work, we consider an EVM below these levels satisfactory.
It follows from the results in Fig.~\ref{fig:bandwidth} that satisfactory EVM is achieved with the ten-tap equalizer for bandwidths up to \SI{72}{MHz} for 16QAM, and up to \SI{96}{MHz} for QPSK.

\footnotetext{The curves with the OFDM label will be discussed in Section \ref{sec:5gnr}.}

\begin{figure}
    \centering
    \includegraphics[width=\linewidth]{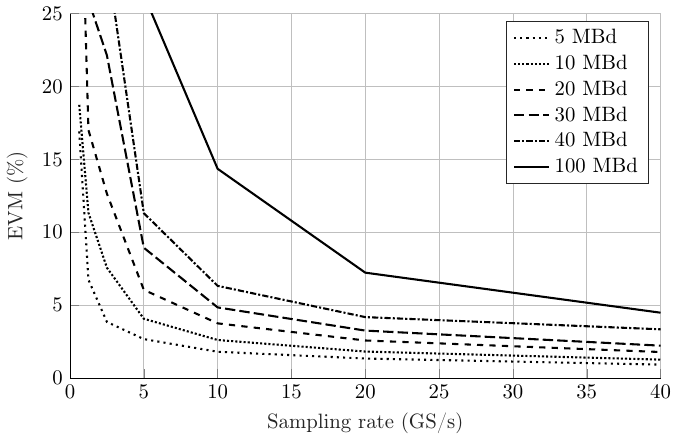}
    \caption{Simulated average per-symbol EVM for different symbol rates and sampling rates.}
    \label{fig:srate}
\end{figure}

In Fig.~\ref{fig:bandwidth} we studied the dependence of the EVM on the signal bandwidth for a fixed sampling rate of \SI{10}{GS/s}.
In Fig.~\ref{fig:srate}, we simulate the impact on the EVM of the sampling rate.
As shown in the figure, increasing the sampling rate beyond \SI{10}{GS/s} is beneficial when the symbol rate is \SI{100}{MBd}, whereas this has only a modest impact on the EVM when the symbol rate is smaller than \SI{40}{MBd}.
In general, as discussed in \cite{Anzhong23}, the choice of the sampling rate involves a trade-off between system performance and fronthaul rate.

\subsubsection{OFDM Waveform}\label{sec:5gnr}
So far, all reported measurements pertain the transmission of a single-carrier signal.
However, such a signal is not relevant for 5G, which uses instead OFDM.
To investigate weather our 1-bit radio-over-fiber fronthaul architecture 
can support OFDM, we perform an over-cable measurement in which an OFDM waveform of \SI{-30}{dBm} average power is fed into the antenna port of an RRH.

In accordance with the NR numerology, we use \SI{60}{kHz} sub-carrier spacing, and a symbol duration (excluding cyclic prefix) of \SI{16.67}{$\mu$s}.
Furthermore, we use least-square channel estimation with a delay-domain window of 20 samples on a Zadoff-Chu pilot sequence.
The pilot sequence is defined by 3GPP in \cite{3gppPilot}.
The power and frequency of the dither signal are not optimized.
Rather, we use the values that were optimal in a similar single-carrier measurement.

In Fig.~\ref{fig:bandwidth}, we present both simulated and measured EVM values as a function of the signal bandwidth.
The EVM curves show a similar trend as in the single-carrier case; however, the EVM values are slightly larger.
This is expected, since the testbed is not optimized for the OFDM waveform.
Despite this, satisfactory EVM is achieved for bandwidths up to \SI{67}{MHz} for 16QAM and up to \SI{87}{MHz} for QPSK.

\subsubsection{Multi-User Reception, In-Band Interference}
\label{sec:mu}
In practical D-MIMO deployment scenarios, the received signal at the RRH may consist of the superposition of signals from different UEs.
The case in which these signals have significantly different power may be problematic for the 1-bit radio-over-fiber architecture, since the dither signal will be optimal only for the stronger among the received signals.

To investigate this scenario, we perform an over-cable measurement where the signal connected to the RRH consists of the superposition of two single-carrier signals at the same frequency, one with fixed power $P_{\text{UE1}}=-30$~\SI{}{dBm}, and one with variable power $P_{\text{UE2}}$.
At the CU, we decode the signal with power $P_{\text{UE1}}$, treating the other signal as noise.

\begin{figure}
    \centering
    \includegraphics[width=\linewidth]{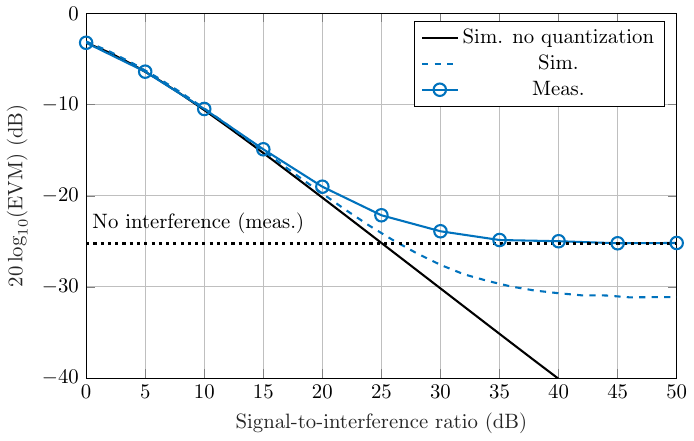}
    \caption{Simulated and measured average per-symbol EVM versus the SIR, for the case of in-band interference.}
    \label{fig:interf}
\end{figure}

In Fig. \ref{fig:interf}, we report the measured and simulated EVM in \SI{}{dB} as a function of SIR $10 \log_{10}(P_{\text{UE1}}/P_{\text{UE2}})$, for a signal with \SI{10}{MBd} symbol rate.
For reference, we include also the simulated EVM for the case of no quantization and infinite SNR, and the measured EVM without interference.

We observe that when the SIR is equal to \SI{15}{dB} or lower, measured and simulated EVM coincide with the simulated EVM for the case of no quantization and infinite SNR.
This implies that, in this regime, the in-band interference dominates over both the thermal and the quantization noise.
As the SIR increases, the measured EVM converges rapidly to the one measured in the absence of inter-symbol interference.
This suggests that the 1-bit radio-over-fiber fronthaul architecture is no more sensitive to this kind of interference than any other architecture.

\subsubsection{Multi-User Reception, Out-of-Band Interference}
Next, we consider the case of out-of-band interference, i.e., interference from a signal transmitted in a bandwidth adjacent to the one of the desired signal.

We consider the transmission of two \SI{4}{MBd} signals centered at \SI{2.3475}{GHz} and \SI{2.3525}{GHz}, respectively.
We fix the power $P_{\text{UE1}}$ of the first signal to \SI{-30}{dBm} and vary the power $P_{\text{UE2}}$ of the second signal.
The dither signal power is optimized for the case in which the SIR is \SI{15}{dB}. 
The dither signal frequency is set to \SI{12}{MHz} and its power to \SI{-6}{dBm}.
At the CU, we decode both signals, each time treating the other signal as noise.

\begin{figure}
    \centering
    \includegraphics[width=\linewidth]{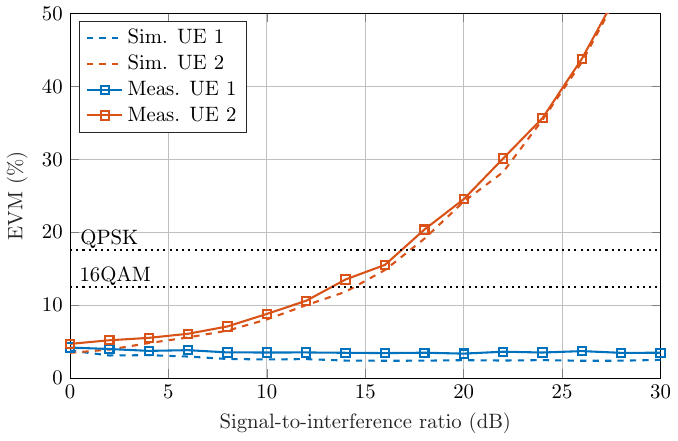}
    \caption{Simulated and measured EVM versus the SIR, for the case of out-of-band interference.}
    \label{fig:fdma}
\end{figure}

In Fig.~\ref{fig:fdma}, we present simulated and measured EVM for both signals, as a function of the SIR.
As expected, when the SIR is \SI{0}{dB}, the two signals have the same EVM since they are subject to the same level of interference.
As the SIR increases, the EVM of the first signal decreases, although slightly, whereas the EVM of the second signal deteriorates.
Note though that 16QAM/QPSK can still be supported even when $P_{\text{UE2}}$ is \SI{13}{dB}/\SI{16}{dB} below $P_{\text{UE1}}$, respectively.

In a D-MIMO setting, where the number of RRHs is large compared to the UEs, each UE is likely to generate a strong signal only for a subset of RRHs.
By requiring that only these RRHs will serve the UE, one can provide reliable service to all UEs in the network.

\section{System Calibration for Over-the-Air Measurements}\label{sec:calibration}
Having confirmed via point-to-point over-cable measurements that low EVM can be achieved, we now consider over-the-air D-MIMO measurements.
In centralized D-MIMO, the uplink channel estimates pertain an effective channel that includes the propagation channel, the RRHs, and the fronthaul. 
A precoding matrix that is based on the uplink channel estimates will then perform poorly if the uplink and downlink hardware are non-reciprocal, i.e., if they affect the signal amplitude and phase differently.
Specifically, the effective baseband uplink channel can be modelled as $\mathbf{H}_{\text{UL}} = \mathbf{R}_{\text{RRH}}\mathbf{H}\mathbf{T}_{\text{UE}}$, where $\mathbf{R}_{\text{RRH}} = \text{diag}(r_1^{\text{RRH}},\dotsc,r_B^{\text{RRH}})$ and $\mathbf{T}_{\text{UE}} = \text{diag}(t_1^{\text{UE}},\dotsc,t_U^{\text{UE}})$ represent the complex-valued gains introduced in the uplink by the $B$ RRHs, the fronthaul, and the $U$ UEs, respectively, and $\mathbf{H} \in \mathbb{C}^{B \times U}$ is the reciprocal channel matrix. 
 Similarly, the effective downlink channel is $\mathbf{H}^{\text{T}}_{\text{DL}} =  \mathbf{R}_{\text{UE}} \mathbf{H}^{\text{T}} \mathbf{T}_{\text{RRH}}$,  where $\mathbf{R}_{\text{UE}} = \text{diag}(r_1^{\text{UE}},\dotsc,r_U^{\text{UE}})$ and $\mathbf{T}_{\text{RRH}} = \text{diag}(t_1^{\text{RRH}},\dotsc,t_B^{\text{RRH}})$ represent the complex-valued gains introduced in the downlink, by the $B$ RRHs, the fronthaul, and the $U$ UEs. 
  
\begin{figure}
    \centering
    \includegraphics[width=\linewidth]{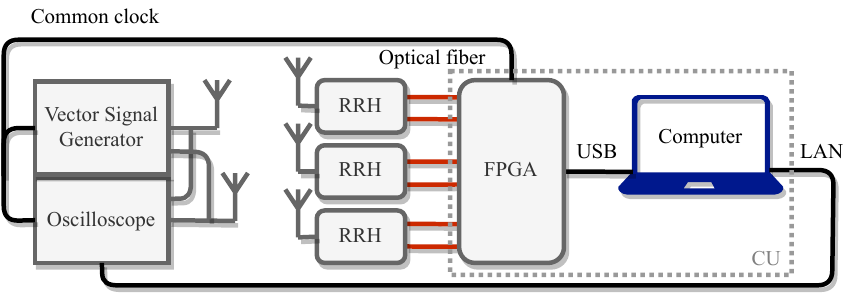}
    \caption{Setup for multi-user D-MIMO over-the-air measurements. The computer is connected via LAN to the vector signal generator and the oscilloscope, which emulates $2$ UE transmitters and receivers, respectively.}
    \label{fig:mumimo}
\end{figure}

 We assess reciprocity by measuring $\mathbf{H}_{\text{UL}}$ and $\mathbf{H}_{\text{DL}}$, using the setup depicted in Fig.~\ref{fig:mumimo}, but for a single UE. 
 The vector signal generator acts as a UE transmitter and the oscilloscope acts as a UE receiver.
Since this measurement equipment emulates ideal transmitter and receiver hardware at the UE, we assume that $\mathbf{T}_{\text{UE}} = \mathbf{R}_{\text{UE}} = \mathbf{I}_U$.
 Furthermore, a common clock is shared between the FPGA, the oscilloscope, and the signal generator, to avoid carrier frequency offsets.
 
\begin{figure}
    \centering
    \includegraphics[width=\linewidth]{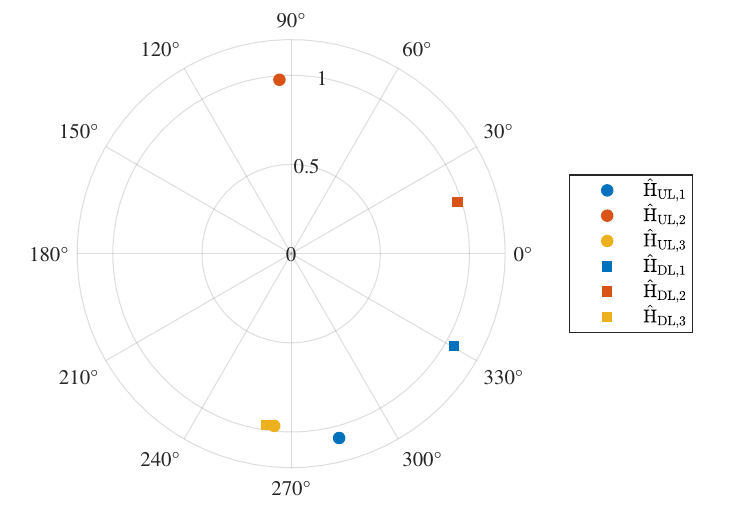}
    \caption{Channel estimates $\widehat{\mathbf{H}}_{\text{UL}} = \begin{bmatrix}\hat{\text{H}}_{\text{UL},1} & \hat{\text{H}}_{\text{UL},2} & \hat{\text{H}}_{\text{UL},3}\end{bmatrix}^{\text{T}}$ (round marks) and $\widehat{\mathbf{H}}_{\text{DL}} = \begin{bmatrix}\hat{\text{H}}_{\text{DL},1} & \hat{\text{H}}_{\text{DL},2} & \hat{\text{H}}_{\text{DL},3}\end{bmatrix}^{\text{T}}$ (square marks) for three RRHs and one UE.}
    \label{fig:ul_vs_dl}
\end{figure}

 We compare the channel estimates $\widehat{\mathbf{H}}_{\text{UL}}$ and $\widehat{\mathbf{H}}_{\text{DL}}$, to assess if reciprocity calibration is required.
 The estimate $\widehat{\mathbf{H}}_{\text{UL}}$ is obtained by orthogonal pilot transmission from the UE to the RRHs, and $\widehat{\mathbf{H}}_{\text{DL}}$ is obtained by orthogonal pilot transmission from the RRHs to the UE.
 The normalized least-square channel estimates are presented in Fig.~\ref{fig:ul_vs_dl}. 
 We observe that the maximum phase difference between two entries corresponding to the same RRH in $\widehat{\mathbf{H}}_{\text{UL}}$ and $\widehat{\mathbf{H}}_{\text{DL}}$ is $75^{\circ}$.
 Hence, the fronthaul is not reciprocal and one needs to perform a reciprocity calibration. 

\begin{figure}
    \centering
    \includegraphics[width=0.85\linewidth]{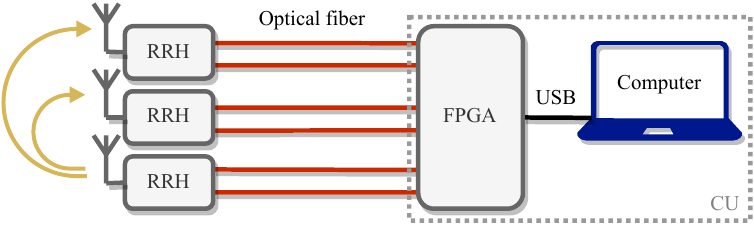}
    \caption{Setup for over-the-air reciprocity calibration. All RRHs exchange signals with each other.}
    \label{fig:cal_lab}
\end{figure}

 To do so, we use the algorithm proposed in \cite{Vieira_Cal17}, in which calibration is achieved by exchanging signals over the air between all RRHs.
 The setup is illustrated in Fig.~\ref{fig:cal_lab}.
 A calibration matrix $\mathbf{C} = \mathbf{T}_{\text{RRH}} \mathbf{R}^{-1}_{\text{RRH}} = \text{diag}(\frac{t_1^{\text{RRH}}}{r_1^{\text{RRH}}},\dotsc,\frac{t_B^{\text{RRH}}}{r_B^{\text{RRH}}})$ is found based on an initial guess of $\mathbf{C}$, followed by an iterative gradient search. 
 Specifically, the cost function $J = \left|\left| \mathbf{y} - \mathbf{R}_{\text{RRH}}\mathbf{H}\mathbf{T_{\text{RRH}}}\mathbf{x} \right|\right|^2 + p$, where $\mathbf{y}$ is the received baseband signal, $\mathbf{x}$ is the transmitted baseband signal, and $p$ is a penalty term, is minimized until a threshold tolerance is fulfilled.
 The precoder based on uplink channel estimates is then multiplied with $\mathbf{C}^{-1}$, to compensate for the lack of reciprocity.
 We discuss the accuracy of the reciprocity calibration in the next section.
 
\section{D-MIMO Over-the-Air Measurements}\label{sec:dmimo}
\begin{figure}
    \centering
    \includegraphics[width=\linewidth]{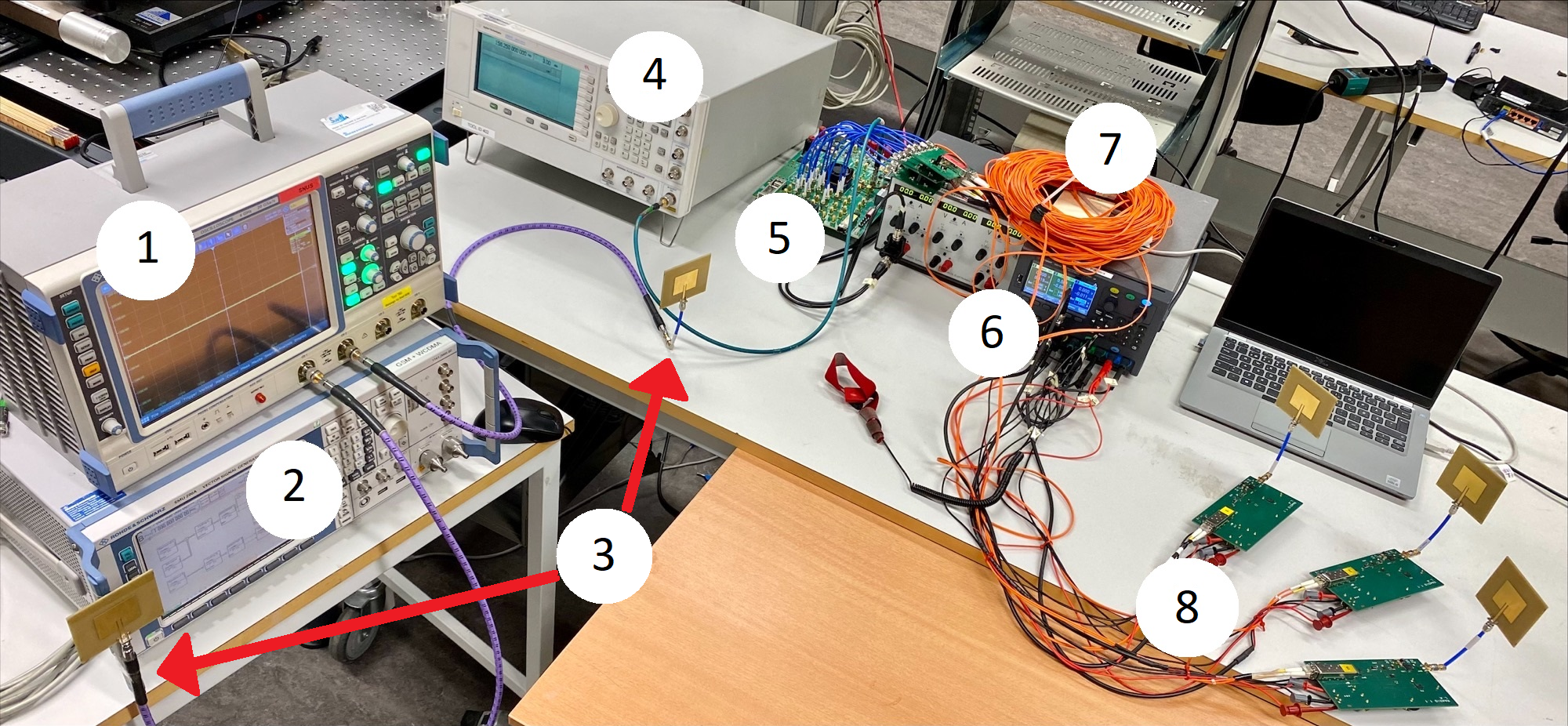}
    \caption{The testbed used for D-MIMO measurements. The instruments are numbered as 1. Oscilloscope, 2. Vector signal generator, 3. UE antennas, 4. Reference oscillator for FPGA, 5. FPGA evaluation board, 6. DC power supplies, 7. Optical fiber cables, 8. RRHs.}
    \label{fig:photo}
\end{figure}

In this section, we present multi-user D-MIMO over-the-air measurements for the case in which $3$ RRHs serve simultaneously $2$ UEs. 
A vector signal generator and an oscilloscope act as the transmitter and receiver for the two UEs, respectively. 
The setup is illustrated in Fig.~\ref{fig:mumimo}, and a picture of the equipment is presented in Fig.~\ref{fig:photo}. 
We consider a line-of-sight scenario, with approximately \SI{2}{m} distance between the UEs and the RRHs.
Furthermore, the two UEs are approximately \SI{1}{m} apart and the distance between the RRHs is \SI{1}{m}.
The measurement distances are limited due to the DC power supply that is shared between all RRHs to control the TDD switch.

\subsection{Downlink}
 We measure the EVM of the received constellation symbols at the UEs when using either $\widehat{\mathbf{H}}_{\text{DL}}$ or $\widehat{\mathbf{H}}_{\text{UL}}$ to create a zero-forcing precoding matrix.
 Specifically, the zero-forcing precoding matrix based on $\widehat{\mathbf{H}}_{\text{DL}}$ is $\mathbf{P}_{\text{DL}}=\widehat{\mathbf{H}}^{*}_{\text{DL}}(\widehat{\mathbf{H}}^{T}_{\text{DL}}\widehat{\mathbf{H}}^{*}_{\text{DL}})^{-1}$.  
 Similarly, the zero-forcing precoding matrix based on $\widehat{\mathbf{H}}_{\text{UL}}$ is $\mathbf{P}_{\text{UL}} = \widehat{\mathbf{H}}^{*}_{\text{UL}}(\widehat{\mathbf{H}}^{T}_{\text{UL}}\widehat{\mathbf{H}}^{*}_{\text{UL}})^{-1}$.
 We also measure the EVM at the UEs when precoding with either $\mathbf{P}_{\text{UL}}$ or $\mathbf{C}^{-1}\mathbf{P}_{\text{UL}}$ to evaluate the impact of reciprocity calibration. 
 
\begin{figure}
\centering
\begin{subfigure}{.5\linewidth}
  \centering
  \includegraphics[height=0.8\linewidth]{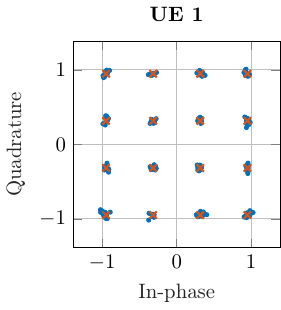}
  \caption{}
  \label{fig:a}
\end{subfigure}%
\begin{subfigure}{.5\linewidth}
  \centering
  \includegraphics[height=0.8\linewidth]{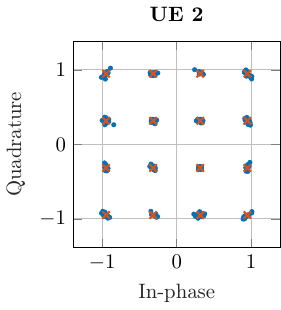}
  \caption{}
  \label{fig:b}
\end{subfigure}
\begin{subfigure}{.5\linewidth}
  \centering
  \includegraphics[height=0.8\linewidth]{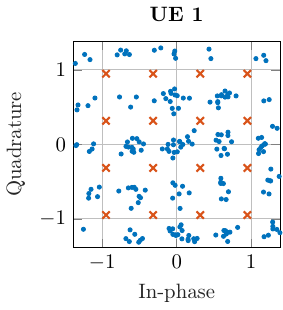}
  \caption{}
  \label{fig:c}
\end{subfigure}%
\begin{subfigure}{.5\linewidth}
  \centering
  \includegraphics[height=0.8\linewidth]{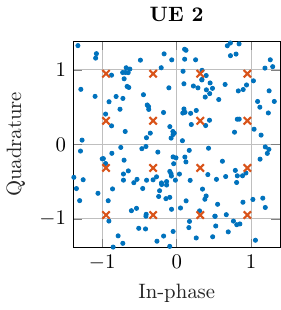}
  \caption{}
  \label{fig:d}
\end{subfigure}
\begin{subfigure}{.5\linewidth}
  \centering
  \includegraphics[height=0.8\linewidth]{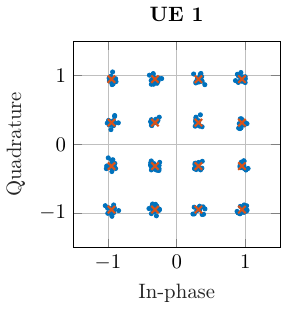}
  \caption{}
  \label{fig:e}
\end{subfigure}%
\begin{subfigure}{.5\linewidth}
  \centering
  \includegraphics[height=0.8\linewidth]{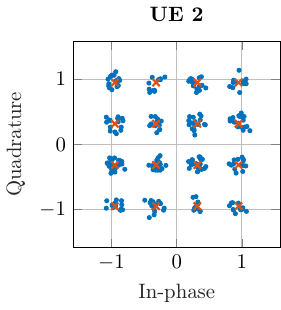}
  \caption{}
  \label{fig:f}
\end{subfigure}
\caption{Constellation diagrams obtained using the precoder based on channel estimation via orthogonal downlink pilots ($\mathbf{P}_{\text{DL}}$), display in \subref{fig:a} $3.7\%$ EVM, and in \subref{fig:b} $3.8\%$ EVM. Constellation diagrams obtained using the precoder based on channel estimation via orthogonal uplink pilots ($\mathbf{P}_{\text{UL}}$), display in \subref{fig:c} $89.8\%$ EVM, and in \subref{fig:d} $84.4\%$ EVM. Constellation diagrams obtained using the precoder based on channel estimation via orthogonal uplink pilots, and calibration is performed ($\mathbf{C}^{-1} \mathbf{P}_{\text{UL}}$), display in \subref{fig:e} $6.4\%$ EVM, and in \subref{fig:f} $10.9\%$ EVM.}
\label{fig:ue_const}
\end{figure}

 The constellation diagram and the EVM for the three different precoders are presented in Fig.~\ref{fig:ue_const}, for the case of a 16QAM signal with \SI{5}{MBd} symbol rate.
 Fig.~\ref{fig:a} and \ref{fig:b} display the received constellation diagrams at the UEs when $\mathbf{P}_{\text{DL}}$ is used for precoding.
 The measured EVM are $3.7\%$ for UE 1 and $3.8\%$ for UE 2, which implies that the two UEs experience similar SNR and that interference is low.
 In Fig.~\ref{fig:c} and \ref{fig:d}, we present the same results for the case in which $\mathbf{P}_{\text{UL}}$ is used for precoding, without reciprocity calibration.
 The measured EVM are $89.8\%$ for UE 1 and $84.4\%$ for UE 2, which is clearly not satisfactory.
 In Fig.~\ref{fig:e} and \ref{fig:f}, $\mathbf{C}^{-1}\mathbf{P}_{\text{UL}}$ is used for precoding.
 The measured EVM are $6.4\%$ for UE 1 and $10.9\%$ for UE 2.
 This confirms that reciprocity calibration is necessary in our setup for TDD operations based on an uplink pilot transmission.
 
 The degradation in EVM when using $\mathbf{C}^{-1} \mathbf{P}_{\text{UL}}$ (Fig. \ref{fig:e} and \ref{fig:f}) compared to $\mathbf{P}_{\text{DL}}$ (Fig. \ref{fig:a} and \ref{fig:b}) for downlink precoding indicates imperfect channel estimation and/or calibration.
 Also, UE $2$ experiences larger EVM than UE $1$, which indicates that the reciprocity calibration can be improved.
 A cause of error is that digital signal processing is performed offline, which involves delays in the order of minutes between when the calibration matrix $\mathbf{C}$ is estimated and when it is used.

\subsection{Uplink}
\begin{figure}
\centering
\begin{subfigure}{.5\linewidth}
  \centering
  \includegraphics[height=0.8\linewidth]{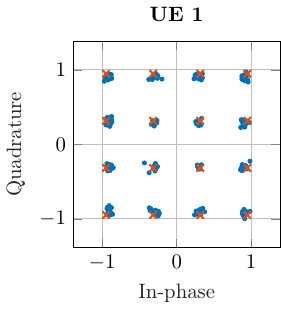}
  \caption{}
  \label{fig:ul_ue1}
\end{subfigure}%
\begin{subfigure}{.5\linewidth}
  \centering
  \includegraphics[height=0.8\linewidth]{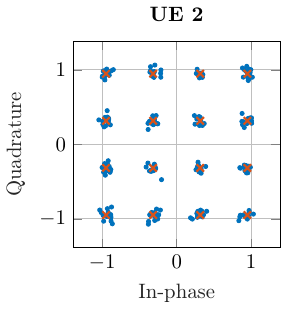}
  \caption{}
  \label{fig:ul_ue2}
\end{subfigure}
\caption{Constellation diagrams for the received signals at the CU displaying in \subref{fig:ul_ue1} $5.2\%$ EVM and in \subref{fig:ul_ue2} $6.9\%$ EVM.}
\label{fig:bs_const}
\end{figure}

We measure the EVM of the received constellation symbols at the CU, after zero-forcing combining.
Specifically, 16QAM modulated signals with \SI{5}{MBd} symbol rate are transmitted from the two UEs simultaneously.
The channel estimate $\widehat{\mathbf{H}}_{\text{UL}}$ is obtained from pilot transmission from the UEs.
The dither frequency is \SI{12}{MHz}, and the power is \SI{-4.5}{dBm}.
In Fig.~\ref{fig:bs_const}, the results are displayed in terms of constellation diagrams. 
The corresponding measured EVM are $5.2\%$ for the signal received from UE 1 and $6.9\%$ for the signal received from UE 2.

We observe that low EVM can be achieved for multi-user uplink transmission, despite the nonlinearity that is introduced by the 1-bit sampling.
Furthermore, we observe that, for this particular measurement scenario, the measured EVM per UE is in the same range as the one measured over cable and presented in Fig. \ref{fig:dynrange} ($5\%$ to $7\%$).
This confirms that the interference between UEs is low and that quantization-unaware spatial processing provides low EVM per UE.

In summary, we conclude that our testbed supports multi-user D-MIMO when equipped with $3$~RRHs serving $2$~UEs.


\section{Discussion}
The experiments conducted in the paper involve just $3$ RRHs.
In practice, a much larger number of RRHs need to be considered to reap the benefits of D-MIMO.
In the following, we present some remarks on the scalability of the proposed architecture.

\begin{itemize}
    \item Number of RRHs: The RRHs in the proposed architecture are based on inexpensive, off-the-shelf components.
    Hence, the proposed architecture remains low-cost even when the number of RRHs increases.
    Since each RRH needs to be connected to the CU via a fiber-optical cable, the number of RRHs that can be deployed is eventually limited by the number of optical ports available at the CU and by the processing speed of the CU.
    This makes the proposed architecture promising for local D-MIMO deployments.

    \item Bandwidth: The sampling rate of \SI{10}{GS/s} of the proposed architecture effectively limits both the carrier frequency and the bandwidth of the transmitted signal.
    In the paper, we have assumed a carrier frequency of \SI{2.35}{GHz} and a bandwidth of up to \SI{100}{MHz}.
    Other carrier frequencies and larger bandwidths can be considered, but this would require one to reoptimize the dither signal and to exchange some of the hardware components at the RRHs (specifically, the BPF and the LPF).
    Also, as shown in Fig.~\ref{fig:srate}, it is beneficial to increase the sampling rate beyond \SI{10}{GS/s} when the bandwidth of the transmitted signal is larger than \SI{40}{MHz}.
    Extensions of the proposed downlink architecture to mm-wave frequencies have been proposed in the literature, and bandwidths up to \SI{700}{MHz} have been considered~\cite{sdm_mmW}.
    Carrier aggregation, which is used primarily in the downlink, can also be supported; indeed, as shown in~\cite{sdm_ca}, $\Sigma\Delta$ modulation is compatible with carrier aggregation: one just needs to equip the RRHs with a set of reconstruction BPFs at the desired central frequencies.

    \item Coverage: The \SI{42}{dB} dynamic range of the AGC at the RRHs determines the coverage of each RRH.
    If we use the indoor office channel model defined in~\cite{3gppChannel} and we assume that the transmit power of the UE is \SI{23}{dBm}, we obtain a maximum non-line-of-sight range of \SI{60}{m}.
    This implies that each RRH will be able to serve only UEs located within this relatively narrow range.
    As a result, TDD switching issues are avoided.
    We would like to emphasize that, using single-mode fiber-optical fronthaul, distances of up to \SI{40}{km} are supported.
    Thus, the proposed optical fronthaul architecture is not the limiting factor in terms of coverage.
\end{itemize}

Measurements for more realistic scenarios, with larger UE--RRH distances, and a larger number of RRHs, will be part of future works.

\section{Conclusion}\label{sec:concl}
We have presented both over-cable and over-the-air uplink and downlink measurements for a D-MIMO testbed operating in TDD mode and involving a 1-bit radio-over-fiber fronthaul.
In the proposed architecture, frequency up- and down-conversion are performed at the CU.
As a consequence, the RRHs do not contain oscillators and mixers, and do not need to be synchronized to enable coherent spatial processing.

We showed that the RRHs can be manufactured using low-cost off-the-shelf components.
Our uplink and downlink measured EVM results for the case of $3$ RRHs serving $2$ UEs show that a satisfactory EVM can be achieved by using linear spatial processing such as zero-forcing, despite the nonlinearity introduced by the 1-bit quantization operation, at the cost of a large oversampling ratio.

We remark that additional RRHs can be added to our measurement setup, with the main limitation being the number of ports available on the FPGA. 
This makes the proposed architecture scalable.

The distance at which the RRHs can be placed in our testbed is currently limited by the TDD switch synchronization, which is implemented by sharing a DC power supply between the RRHs.
As part of future work, we plan to control each switch via a data-channel cable between the CU and the RRH, to be deployed together with the optical fiber, as part of the fronthaul.


\ifCLASSOPTIONcaptionsoff
  \newpage
\fi

\typeout{}
\bibliographystyle{IEEEtran}
\bibliography{extracted}

\begin{thebibliography}{10}
\providecommand{\url}[1]{#1}
\csname url@samestyle\endcsname
\providecommand{\newblock}{\relax}
\providecommand{\bibinfo}[2]{#2}
\providecommand{\BIBentrySTDinterwordspacing}{\spaceskip=0pt\relax}
\providecommand{\BIBentryALTinterwordstretchfactor}{4}
\providecommand{\BIBentryALTinterwordspacing}{\spaceskip=\fontdimen2\font plus
\BIBentryALTinterwordstretchfactor\fontdimen3\font minus \fontdimen4\font\relax}
\providecommand{\BIBforeignlanguage}[2]{{%
\expandafter\ifx\csname l@#1\endcsname\relax
\typeout{** WARNING: IEEEtran.bst: No hyphenation pattern has been}%
\typeout{** loaded for the language `#1'. Using the pattern for}%
\typeout{** the default language instead.}%
\else
\language=\csname l@#1\endcsname
\fi
#2}}
\providecommand{\BIBdecl}{\relax}
\BIBdecl

\bibitem{aabel_asilomar20}
L.~Aabel, G.~Durisi, I.~C. Sezgin, S.~Jacobsson, C.~Fager, and M.~Coldrey, ``Distributed massive {MIMO} via all-digital radio over fiber,'' in \emph{Proc. Asilomar Conf. Signals, Syst., Comput.}, Pacific Grove, CA, USA, Jun. 2020.

\bibitem{Prata2016All-digitalC-RAN}
A.~Prata, A.~S. Oliveira, and N.~B. Carvalho, ``All-digital flexible uplink remote radio head for {C-RAN},'' in \emph{{IEEE} {MTTS} Int. Microw. Symp.}, San Francisco, CA, USA, May 2016, pp. 1--4.

\bibitem{IbraMe21}
I.~C. Sezgin, L.~Aabel, S.~Jacobsson, G.~Durisi, Z.~S. He, and C.~Fager, ``All-digital, radio-over-fiber, communication link architecture for time-division duplex distributed antenna systems,'' \emph{J. Lightw. Technol.}, vol.~39, no.~9, pp. 2769--2779, May 2021.

\bibitem{Maier2011ClassO}
S.~Maier, W.~Kuebart, C.~Haslach, U.~Seyfried, W.~Templ, A.~Frotzcher, D.~Markert, R.~Matz, and A.~Pascht, ``Class-o base station system with {RF} pulse-width-modulation in downlink and uplink,'' in \emph{Asia-Pacific Microw. Conf.}, Melbourne, VIC, Australia, Dec. 2011, pp. 1222--1225.

\bibitem{ulf_6g21}
U.~Gustavsson, P.~Frenger, C.~Fager, T.~Eriksson, H.~Zirath, F.~Dielacher, C.~Studer, A.~P{\"a}rssinen, R.~Correia, J.~N. Matos, D.~Belo, and N.~B. Carvalho, ``Implementation challenges and opportunities in {beyond-5G} and {6G} communication,'' \emph{{IEEE} J. Microwaves.}, vol.~1, no.~1, pp. 86--100, Jan. 2021.

\bibitem{hexaX_23}
B.~M. Khorsandi \emph{et~al.}, ``{Hexa-X} architecture for {B5G/6G} networks,'' Hexa-X, Tech. Rep., July 2023.

\bibitem{zhou_dmimo03}
S.~Zhou, M.~Zhao, X.~Xu, J.~Wang, and Y.~Yao, ``Distributed wireless communication system: A new architecture for future public wireless access,'' \emph{{IEEE} Commun. Mag.}, vol.~41, no.~3, pp. 108--113, Mar. 2003.

\bibitem{Ngo2017Cell-FreeCells}
H.~Q. {Ngo}, A.~{Ashikhmin}, H.~{Yang}, E.~G. {Larsson}, and T.~L. {Marzetta}, ``Cell-free massive {MIMO} versus small cells,'' \emph{{IEEE} Trans. Commun.}, vol.~16, no.~3, pp. 1834--1850, Mar. 2017.

\bibitem{bjornson_cellfree}
E.~Bj{\"o}rnson and L.~Sanguinetti, ``Making cell-free massive {MIMO} competitive with {MMSE} processing and centralized implementation,'' \emph{{IEEE} Trans. Wireless Commun.}, vol.~19, no.~1, pp. 77--90, Jan. 2020.

\bibitem{cellfreebook}
{\"O}.~T. Demir, E.~Bj{\"o}rnson, and L.~Sanguinetti, \emph{{Foundations of user-centric cell-free massive MIMO}}.\hskip 1em plus 0.5em minus 0.4em\relax Now Publisher, Foundations and Trends in Signal Processing, 2021.

\bibitem{FridaConf}
F.~{Olofsson}, L.~{Aabel}, M.~{Karlsson}, and C.~{Fager}, ``Comparison of transmitter nonlinearity impairments in externally modulated sigma-delta-over-fiber vs analog radio-over-fiber links,'' in \emph{Opt. Fiber Commun. Conf. Exhibit. (OFC)}, San Diego, CA, USA, Mar. 2022.

\bibitem{puerta_arof}
R.~Puerta, M.~Han, M.~Joharifar, R.~Schatz, Y.-T. Sun, Y.~Fan, A.~Djupsjöbacka, G.~Maisons, J.~Abautret, R.~Teissier, L.~Zhang, S.~Spolitis, M.~Wang, V.~Bobrovs, S.~Lourdudoss, X.~Yu, S.~Popov, O.~Ozolins, and X.~Pang, ``{NR} conformance testing of analog radio-over-{LWIR} {FSO} fronthaul link for {6G} distributed {MIMO} networks,'' in \emph{Opt. Fiber Commun. Conf. Exhibit. {(OFC)}}, San Diego, CA, USA, Mar. 2023.

\bibitem{Sezgin2018SISO}
I.~C. {Sezgin}, T.~{Eriksson}, J.~{Gustavsson}, and C.~{Fager}, ``A flexible multi-{Gbps} transmitter using ultra-high speed sigma-delta-over-fiber,'' in \emph{{IEEE} {MTTS} Int. Microw. Symp.}, Philadelphia, PA, USA, Jun. 2018, pp. 1195--1198.

\bibitem{Sezgin2019MIMOtestbed}
I.~C. Sezgin, M.~Dahlgren, T.~Eriksson, M.~Coldrey, C.~Larsson, J.~Gustavsson, and C.~Fager, ``A low-complexity distributed-{MIMO} testbed based on high-speed sigma-delta-over-fiber,'' \emph{{IEEE} Trans. Microw. Theory Techn.}, vol.~67, no.~7, pp. 2861--2872, Jul. 2019.

\bibitem{Sezgin2019MIMOeval}
I.~C. {Sezgin}, T.~{Eriksson}, J.~{Gustavsson}, and C.~{Fager}, ``Evaluation of distributed {MIMO} communication using a low-complexity sigma-delta-over-fiber testbed,'' in \emph{{IEEE} {MTTS} Int. Microw. Symp.}, Boston, MA, USA, Jun. 2019, pp. 754--757.

\bibitem{Wang2019}
J.~Wang, Z.~Jia, L.~A. Campos, and C.~Knittle, ``Delta-sigma modulation for next generation fronthaul interface,'' \emph{J. Lightw. Technol.}, vol.~37, no.~12, pp. 2838--2850, Jun. 2019.

\bibitem{Wu_rfsdm20}
C.~Wu, H.~Li, O.~Caytan, J.~{Van Kerrebrouck}, L.~Breyne, J.~Bauwelinck, P.~Demeester, and G.~Torfs, ``Distributed multi-user {MIMO} transmission using real-time sigma-delta-over-fiber for next generation fronthaul interface,'' \emph{J. Lightw. Technol.}, vol.~38, no.~4, pp. 4229--4240, Feb. 2020.

\bibitem{JacobssonGlobecom}
S.~{Jacobsson}, L.~{Aabel}, M.~{Coldrey}, I.~C. {Sezgin}, C.~{Fager}, G.~{Durisi}, and C.~{Studer}, ``Massive {MU-MIMO-OFDM} uplink with direct {RF}-sampling and 1-bit {ADCs},'' in \emph{Proc. IEEE Global Telecommun. Conf. (GLOBECOM)}, Waikoloa, HI, USA, Dec. 2019.

\bibitem{Anzhong23}
A.~Hu, L.~Aabel, G.~Durisi, S.~Jacobsson, M.~Coldrey, C.~Fager, and C.~Studer, ``{EVM} analysis of distributed massive {MIMO} with 1-bit radio-over-fiber fronthaul,'' 2023, manuscript submitted for publication.

\bibitem{Pavan_sdmBook17}
S.~Pavan, R.~Schreier, and G.~C. Temes, \emph{{Understanding Delta-Sigma Data Converters}}, 2nd~ed., R.~Jacob~Baker, Ed.\hskip 1em plus 0.5em minus 0.4em\relax Hoboken, NJ, USA: John Wiley {\&} Sons, 2017.

\bibitem{Cordeiro2017}
R.~F. Cordeiro, A.~Prata, A.~S. Oliveira, J.~M. Vieira, and N.~B. De~Carvalho, ``Agile all-digital {RF} transceiver implemented in {FPGA},'' \emph{{IEEE} Trans. Microw. Theory Techn.}, vol.~65, no.~11, pp. 4229--4240, Nov. 2017.

\bibitem{Mouton_pwm14}
H.~du~Toit~Mouton, B.~McGrath, D.~Grahame~Holmes, and R.~H. Wilkinson, ``One-dimensional spectral analysis of complex pwm waveforms using superposition,'' \emph{{IEEE} Trans. Power Electron.}, vol.~29, no.~12, pp. 6762--6778, Dec. 2014.

\bibitem{fpga}
{Altera Corporation}, ``Transceiver signal integrity development kit, {Stratix} {V} {GT} edition reference manual,'' Jan. 2016.

\bibitem{sfp}
{Avago Technologies}, ``Fiber optical transceiver ({AFBR-709SMZ}),'' Jan. 2013.

\bibitem{3gpp}
{3GPP}, ``{Evolved universal terrestrial radio access (E-UTRA); base station (BS) radio transmission and reception},'' 2022, {TS 36.104 version 17.7.0 Rel. 17}.

\bibitem{5g}
------, ``{NR; base station (BS) radio transmission and reception},'' 2022, {TS 38.104 version 17.7.0 Rel. 17}.

\bibitem{3gppPilot}
------, ``{5G; NR; Physical channels and modulation},'' 2020, {TS 38.211 version 16.2.0 Rel. 16}.

\bibitem{Vieira_Cal17}
J.~Vieira, F.~Rusek, O.~Edfors, S.~Malkowsky, L.~Liu, and F.~Tufvesson, ``Reciprocity calibration for massive {MIMO}: Proposal, modeling, and validation,'' \emph{{IEEE} Trans. Wireless Commun.}, vol.~16, no.~5, pp. 3042--3056, May 2017.

\bibitem{sdm_mmW}
H.~Bao, F.~Ponzini, and C.~Fager, ``Flexible mm-wave sigma-delta-over-fiber {MIMO} link,'' \emph{J. Lightw. Technol.}, vol.~41, no.~14, pp. 4734--4742, 2023.

\bibitem{sdm_ca}
N.~V. Silva, A.~S.~R. Oliveira, U.~Gustavsson, and N.~B. Carvalho, ``A novel all-digital multichannel multimode {RF} transmitter using delta-sigma modulation,'' \emph{{IEEE} Microw. and Wirel. Compon. Lett.}, vol.~12, no.~3, pp. 156--158, Mar. 2012.

\bibitem{3gppChannel}
{3GPP}, ``{Technical Specification Group Radio Access Network; Study on channel model for frequencies from 0.5 to 100 GHz},'' 2023, {TS 38.109 version 17.1.0 Rel. 17}.

\end{thebibliography}

\vfill

\end{document}